\documentclass[a4paper,fleqn,usenatbib]{mnras}

\usepackage[T1]{fontenc}
\usepackage{ae,aecompl}
\usepackage[normalem]{ulem}

\usepackage{graphicx}	
\usepackage{amsmath}	
\usepackage{amssymb}	

\newcommand{\PSRnumPKSFAST}{11 } 
\newcommand{\PSRnumBACKGROUNDppdot}{2217 } 
\newcommand{\PSRnumBACKGROUNDrmdm}{1166 } 
\newcommand{\PSRnumFASTcands}{70 } 
\newcommand{\PSRnumFASTcandsRECYC}{3 } 
\newcommand{\PSRnumFASTcandsPKS}{28 } 
\newcommand{\PSRnumFASTconfPKS}{13 } 
\newcommand{\PSRnumFASTnonconfPKS}{14 } 
\newcommand{\PSRnumFASTobsPKS}{27 } 

\title[Discovery of \PSRnumPKSFAST pulsars by FAST]{An in-depth investigation of \PSRnumPKSFAST pulsars discovered by FAST.}

\author[A. D. Cameron et al.]{
A.~D.~Cameron,$^{1,2}$\thanks{E-mail: andrew.cameron@csiro.au} 
D.~Li$^{1,3}$\thanks{E-mail: dili@nao.cas.cn},
G.~Hobbs$^{2,1}$,
L.~Zhang$^{1,2,3}$, 
C.~C.~Miao$^{1,3}$,\newauthor
J.~B.~Wang$^{4,5,6}$, 
M.~Yuan$^{1,3}$,
S.~Wang$^{1,3}$,
G.~Jacobs Corban$^{2,7}$,
M.~Cruces$^{8}$,
S.~Dai$^{1,2}$, \newauthor
Y.~Feng$^{1,2,3}$,
J.~Han$^{1,3}$,
J.~F.~Kaczmarek$^{2,9}$, 
J.~R.~Niu$^{1,3}$,
Z.~C.~Pan$^{1}$, 
L.~Qian$^{1}$,\newauthor
Z.~Z.~Tao$^{1,10,11}$,
P.~Wang$^{1}$,
S.~Q.~Wang$^{2,3,4}$,
H.~Xu$^{1,3}$,
R.~X.~Xu$^{12,13}$,
Y.~L.~Yue$^{1}$,\newauthor
S.~B.~Zhang$^{2,3,14,15}$, 
Q.~J.~Zhi$^{10,11}$,
W.~W.~Zhu$^{1}$, 
D.~J.~Champion$^{8}$,
M.~Kramer$^{8,16}$,\newauthor
S.~Q.~Zhou$^{1,3}$,
K.~P.~Qiu$^{17}$,
M.~Zhu$^{1}$.
\\
$^1$ National Astronomical Observatories, Chinese Academy of Sciences, 20A Datun Road, Chaoyang District, Beijing 100101, China \\ 
$^2$ CSIRO Astronomy and Space Science, PO Box 76, Epping, NSW 1710, Australia\\
$^3$ University of Chinese Academy of Sciences, Beijing 100049, China\\
$^4$ Xinjiang Astronomical Observatory, 150, Science-1 Street, Urumqi, 830011 Xinjiang, China \\
$^{5}$ Key Laboratory of Radio Astronomy, Chinese Academy of Sciences, 150 Science 1-Street, Urumqi, 830011 Xinjiang, China \\
$^{6}$ Xinjiang Key Laboratory of Radio Astrophysics, 150 Science 1-Street, Urumqi, 830011 Xinjiang, China \\
$^{7}$ Victoria University of Wellington, PO Box 600, Wellington 6140, New Zealand\\
$^{8}$ Max-Planck Institut f{\"u}r Radioastronomie, Auf dem H{\"u}gel 69, D-53121 Bonn, Germany\\
$^{9}$ National Research Council Canada, Herzberg Research Centre for Astronomy and Astrophysics, Dominion Radio Astrophysical Observatory, P.O. Box 248, Penticton, British Columbia, V2A 6J9, Canada\\
$^{10}$ Guizhou Provincial Key Laboratory of Radio Astronomy and Data Processing, Guizhou Normal University, Guiyang 550001, China \\
$^{11}$ School of Physics and Electronic Science, Guizhou Normal University, Guiyang 550001, China \\
$^{12}$ Kavli Institute for Astronomy and Astrophysics, Peking University, Beijing 100871, China \\
$^{13}$ Department of Astronomy, School of Physics, Peking University, Beijing 100871, China \\
$^{14}$ Purple Mountain Observatory, Chinese Academy of Sciences, Nanjing 210008, China \\
$^{15}$ International Centre for Radio Astronomy Research, University of Western Australia, Crawley, WA 6009, Australia\\
$^{16}$ Jodrell Bank Center for Astrophysics, University of Manchester, Alan Turing Building, Oxford Road, Manchester M13 9PL, United Kingdom \\
$^{17}$ School of Astronomy and Space Science, Nanjing University, 163 Xianlin Avenue, Nanjing 210023, China\\
}

\date{Accepted XXX. Received YYY; in original form ZZZ}

\pubyear{2020}

\begin{document}
\label{firstpage}
\pagerange{\pageref{firstpage}--\pageref{lastpage}}
\maketitle

\begin{abstract}
We present timing solutions and analyses of \PSRnumPKSFAST pulsars discovered by the Five-hundred-meter Aperture Spherical radio Telescope (FAST). These pulsars were discovered using an ultra-wide bandwidth receiver in drift-scan observations made during the commissioning phase of FAST, and were then confirmed and timed using the 64-m Parkes Radio Telescope. Each pulsar has been observed over a span of at least one year. Highlighted discoveries include PSR~J0344$-$0901, which displays mode-changing behaviour and may belong to the class of so-called `swooshing' pulsars (alongside PSRs~B0919$+$06 and B1859$+$07); PSR~J0803$-$0942, whose emission is almost completely linearly polarised; and PSRs~J1900$-$0134 and J1945$+$1211, whose well defined polarisation angle curves place stringent constraints on their emission geometry. We further discuss the detectability of these pulsars by earlier surveys, and highlight lessons learned from our work in carrying out confirmation and monitoring observations of pulsars discovered by a highly sensitive telescope, many of which may be applicable to next-generation pulsar surveys. This paper marks one of the first major releases of FAST-discovered pulsars, and paves the way for future discoveries anticipated from the Commensal Radio Astronomy FAST Survey (CRAFTS).
\end{abstract}




\section{Introduction}\label{sec: introduction}

The completion of the Five-hundred-meter Aperture Spherical radio Telescope \citep[FAST;][]{nlj+11,li16fast,jth+20}, located in China's Guizhou province, represents a significant achievement in radio astronomy. FAST currently stands as the world's largest single-dish radio telescope, and the most sensitive operating within the 270 to 1620\,MHz band. This increase in sensitivity naturally promises new advances in a number of areas of research \citep[e.g.\ see the mini-volume published on Research in Astronomy \& Astrophysics;][]{li19raa}. The discovery and study of radio pulsars, a mainstay of single-dish radio astronomy for over 50 years, is one area of particularly high priority.

Between August~2017 and February~2018, FAST conducted an initial set of driftscan observations in a pulsar search mode using its ultra-wide-bandwidth receiver (UWB). While these observations were primarily for commissioning purposes, they also served as a pathfinder for the Commensal Radio Astronomy FAST Survey \citep[CRAFTS;][]{lwq+18}, allowing for the development of pulsar search pipelines and computational infrastructure, which will eventually be applied to the full-scale survey. CRAFTS, along with the Search of Pulsars in Special Population \citep[SP$^2$;][]{prl+20} represent the two major pulsar search efforts to be carried out by FAST.

To date, \PSRnumFASTcands candidates have been identified using the UWB, out of which 51 have been confirmed\footnote{The list of confirmed new pulsars can be found at \url{https://crafts.bao.ac.cn}} in this initial set of drift-scan observations. FAST, however, is limited in its efficiency at both confirming and timing these new pulsars. This was largely due to the fact that FAST's slew rate is relatively slow, with a regular source changing time between 5 to 10 minutes. This makes short follow-up observations of multiple targets, the most common mode of regular timing, less efficient. Therefore, the CRAFTS team elected to purchase observing time on the 64-m Parkes Radio Telescope to facilitate these follow-up observations. As Parkes does not have a perfect overlap in sky coverage with FAST, the 100-m Effelsberg Radio Telescope was also used in order to follow-up those pulsars in the northern portion of the FAST sky (Cruces et~al. in prep.).

In this paper we describe \PSRnumPKSFAST pulsars, discovered by FAST with the UWB receiver, that have now been monitored with the Parkes telescope for at least one year, thereby enabling a timing solution for each pulsar to be determined. In Section~\ref{sec: methodology} we describe the Parkes observations along with our confirmation and timing strategy. The timing solutions for each of the \PSRnumPKSFAST pulsars presented in this paper are listed in Section~\ref{sec: properties}, while Section~\ref{sec: pulsars of interest} describes the unique features of interest of each pulsar in turn. In Section~\ref{sec: discussion} we discuss both whether FAST was necessary in the discovery of these pulsars, and why more FAST pulsars were not confirmed by Parkes despite attempts to do so. We also highlight lessons learned as part of this project which may be applicable to the next generation of pulsar surveys. We finally summarise our conclusions in Section~\ref{sec: conclusions}.

\section{Observations \& Data Reduction}\label{sec: methodology}

 The UWB is a single-pixel receiver observing between frequencies of 270 to 1620\,MHz. The system temperature $T_\text{sys}$ of the receiver was measured as being between $60\,\text{K}<T_\text{sys}<70\,\text{K}$ \citep{qpl+19}. The full width at half maximum (FWHM) of the UWB beam at 270\,MHz is about 15\arcmin, corresponding to a source crossing time of one minute at $\delta=0^{\circ}$.
 
Each observation recorded by FAST was searched independently by three data-processing pipelines, incorporating both periodicity and single-pulsar search techniques. Due to the impact of radio-frequency interference (RFI) and the dropping efficiency of the UWB beyond 1GHz, the searches were primarily carried out in radio frequencies below $\sim800\,\text{MHz}$. Further details of these observations and the applied pulsar-searching techniques can be found in \cite{qpl+19} and \cite{ypq+19}.

A total of \PSRnumFASTcands pulsar candidates have been identified in data recorded by FAST's UWB as part of its driftscan observations. Of the \PSRnumFASTcands wideband candidates, only \PSRnumFASTcandsPKS are visible from the Parkes Radio Telescope. Confirmation observations of these candidates with Parkes began in late 2017 using the 21-cm multibeam receiver \citep[MB20;][]{swb+96} in combination with multiple backend systems, including the Berkeley Parkes Swinburne Recorder\footnote{http://www.astronomy.swin.edu.au/pulsar/?topic=bpsr} (BPSR) and the Parkes Digital Filter Bank backend system (PDFB4). The observing specifications of both of these configurations can be found in Table~\ref{tab: observing setup}.

These confirmation observations were conducted on an ad-hoc and opportunistic basis in order to maximise the number of confirmed pulsars within only a limited amount of available observing time. As a result, we did not implement a formal protocol when attempting to follow-up each candidate. An initial confirmation observation typically consisted of an initial 3600-s long scan centered on the candidate position as provided by the FAST team. Each observation was then processed through a \textsc{presto}\footnote{http://www.cv.nrao.edu/$\sim$sransom/presto}-based periodicity search \citep{ransom01}, including RFI-mitigation in both the time and frequency domains as well as a limited acceleration search (typically as high as $\sim5\,\text{m\,s}^{-2}$). In the event of a non-detection of the candidate, further observations were conducted based upon the merits of the candidate signal. These included repeated observations on the central position (to account for any scintillation effects) with scans up to $\sim7000\,\text{s}$ in duration, as well as offset-gridding observations in an attempt to account for errors in the provided candidate positions. These grids typically followed the `ring-of-3' technique set out in \cite{ncb15}.

In summary, of the \PSRnumFASTcandsPKS pulsar candidates visible from Parkes, \PSRnumFASTobsPKS candidates were observed, resulting in \PSRnumFASTconfPKS successful pulsar confirmations. Of these \PSRnumFASTconfPKS pulsars, one (PSR~J1822$+$26) was handed over for long-term timing and monitoring by Effelsberg due to its high northern declination and will feature in a separate publication (Cruces et~al. in prep.). Another (PSR~J1824$-$0132) was found to have already been discovered as part of the High Time Resolution Universe Southern Intermediate Latitude Survey \citep[HTRU-S MedLat;][]{bsb+19}, and so was not pursued for further study. A further pulsar (PSR~J1851$-$0633) was found to have been previously been discovered by the HTRU-S Galactic Plane Survey \citep[HTRU-S LowLat;][]{ccb+20}, but lacked a timing solution\footnote{PSR~J1851$-$0633 is listed in \cite{ccb+20}. as PSR~J1851$-$06.} and was therefore retained for further study. After accounting for these caveats, \PSRnumPKSFAST confirmed pulsars remained as part of our long term-monitoring campaign.

Once a pulsar candidate was confirmed by Parkes, it was then assigned to a long-term monitoring program in order to determine a phase-connected timing solution and further explore its emission properties. Each pulsar was observed with an irregular cadence, with observations typically separated by $1-2$ weeks. These were initially carried out using the MB20 receiver and PDFB4 backend, using the configuration reported in Table~\ref{tab: observing setup}. However, due to the role of Parkes in tracking the passage of the Voyager 2 spacecraft through the Solar System's heliopause region \citep[see e.g.][]{voyager2}, the MB20 receiver was removed from the telescope in mid-October of 2018. Monitoring observations were then transitioned to the recently-commissioned ultra-wide-bandwidth low-frequency (UWL) receiver \citep{hmd+19}, which can observe over a continuous bandwidth between 704 and 4032\,MHz. However, although wideband data was recorded simultaneously using the MEDUSA backend for all UWL observations, this paper will only consider the narrow-band PDFB4 data (256\,MHz centered at 1369\,MHz). A study of the wideband properties of each of the \PSRnumPKSFAST pulsars discussed in this paper will be reserved for a future publication.

\begin{table*}
\caption{Parameters for the observations conducted as part of this research, including the system temperature (\protect$T_\text{sys}$), central frequency (\protect$f_\text{c}$), bandwidth (\protect$\Delta f$), number of frequency channels (\protect$n_\text{chan}$), and the coherent de-dispersion capability of each receiver/backend combination. Parameters for the UWL are taken from Table~2 of \protect\cite{hmd+19}.}\label{tab: observing setup}
\begin{center}
\begin{tabular}{lllllll}
\hline
Receiver & Backend & Coherent & $T_\text{sys}$ & $f_\text{c}$ & $\Delta f$ & $n_\text{chan}$ \\
 & & & (K) & (MHz) & (MHz) & \\
\hline
MB20 & PDFB4 & N & 23 & 1369 & 256 & 1024$^{\text{a}}$\\
 & BPSR & N & 23 & 1382 & 400$^{\text{b}}$ & 1024\\
 \hline
UWL & PDFB4 & N & 23 & 1369 & 256 & 1024$^{\text{a}}$\\
\hline
\multicolumn{7}{l}{\footnotesize{$^{\text{a}}$ PDFB4 search mode observations were recorded with 512 channels.}} \\
\multicolumn{7}{l}{\footnotesize{$^{\text{b}}$ The usable bandwidth of BPSR is reduced to 340\,MHz due to}} \\
  \multicolumn{7}{l}{\footnotesize{strong RFI from the \textit{Thuraya 3} satellite above $\sim$1540\,MHz.}} \\
\end{tabular}
\end{center}
\end{table*}

All observations\footnote{Observations for this paper were recorded under Parkes project codes PX500 and PX501, with additional data for PSR~J1926$-$0652 recorded under P863 and P985.} recorded by both the MB20 and UWL receivers were processed through a standard reduction methodology. Each \textsc{psrfits}-format timing archive was manually cleaned in both the time and frequency domains following visual inspection using the \textsc{psrchive}\footnote{http://psrchive.sourceforge.net} software package \citep{hvsm04}. Polarisation calibration was conducted against an observation of a pulsed noise diode in order to account for the differential gain and phase between the receiver's polarisation feeds. Each observation was then flux calibrated with reference to observations taken of Hydra~A\footnote{Recorded as part of the Parkes Pulsar Timing Array (PPTA) project \citep[see e.g.][]{mhb+13,xwh+19}.}, which were typically separated in time from each pulsar observation by one to two weeks.

\section{Pulsar properties and analysis}\label{sec: properties}

Timing solutions for each of the \PSRnumPKSFAST pulsars are given in Tables~\ref{tab: timing table 1}, \ref{tab: timing table 2} and \ref{tab: timing table 3}. We note that both PSR~J1900$-$0134 \citep{qpl+19} and PSR~J1926$-$0652 \citep{zlh+19} have previously been published with complete timing solutions based on earlier versions of the data presented here. In each case, our updated solution represents a significant improvement in timing precision. 

\begin{figure}
\begin{center}
\includegraphics[height=\columnwidth]{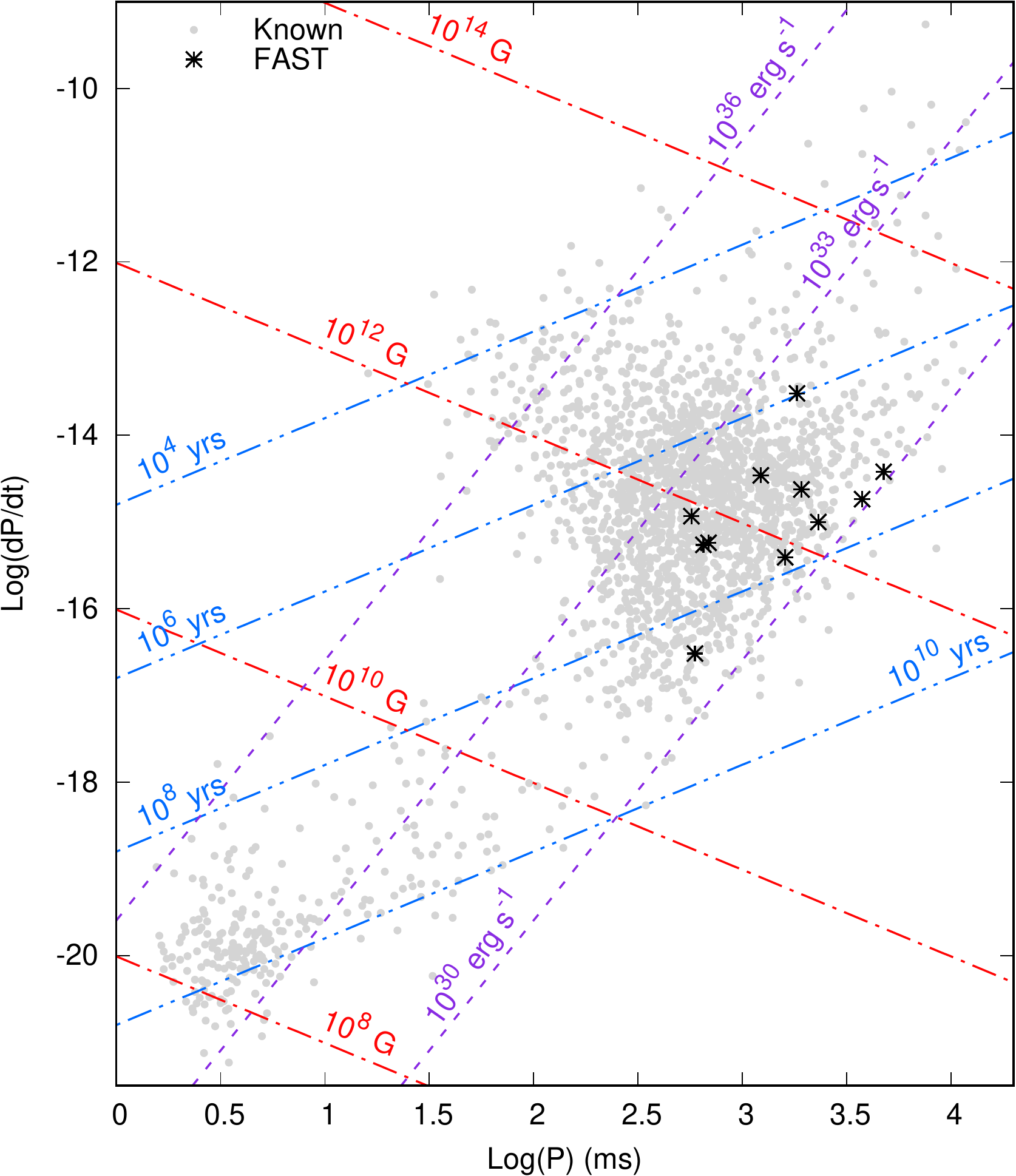} 
\end{center}
\caption{A $P$-$\dot{P}$ diagram displaying the \PSRnumPKSFAST newly-discovered pulsars against the known pulsar population, consisting of \PSRnumBACKGROUNDppdot  pulsars for which \textsc{psrcat} \citep{psrcat} lists well-measured values of $P$ and $\dot{P}$. Also shown are lines of constant spin-down luminosity (purple, single dashes), surface magnetic field strength (red, dash-dot) and characteristic age (blue, dash-double dot). Note that PSRs~J0529$-$0715 and J1919$+$2621 appear almost co-incident in this plot, due to their very similar values of $P$ and $\dot{P}$.}\label{fig: discovery p-pdot} 
\end{figure}

Figure~\ref{fig: discovery p-pdot} shows our \PSRnumPKSFAST pulsars plotted on a $P$-$\dot{P}$ diagram against the known pulsar population. All \PSRnumPKSFAST pulsars are situated within the island of canonical, unrecycled pulsars, and appear to represent a collection of older, slower pulsars. Although it is hard to draw meaningful conclusions based on such a small collection of pulsars, it is notable that only \PSRnumFASTcandsRECYC of the \PSRnumFASTcands candidates identified by FAST's UWB receiver have periods that would potentially qualify them as recycled millisecond pulsars (MSPs). The \PSRnumPKSFAST pulsars we report would therefore seem to be representative of the larger set of FAST UWB candidates. It is likely that this apparent population bias is due to a number of overlapping selection effects.

In order to develop each pulsar's timing solution, each observation was processed through a range of software packages including \textsc{psrchive}, \textsc{dspsr}\footnote{https://sourceforge.net/projects/dspsr} \citep{dspsr}, \textsc{sigproc}\footnote{http://sigproc.sourceforge.net} \citep{sigproc}, \textsc{presto}, \textsc{tempo}\footnote{http://tempo.sourceforge.net} and \textsc{tempo2}\footnote{http://www.atnf.csiro.au/research/pulsar/tempo2} \citep{hem06}. Pulse times-of-arrival (TOAs) were first produced by summing each observation in both frequency and polarisation, before partially summing in time and cross-correlating each summed profile against a standard reference pulse profile. Initial timing solutions were then developed using \textsc{tempo} together with a modified version of the \textsc{dracula}\footnote{https://github.com/pfreire163/Dracula} software package, which solves for the global rotation count of a pulsar between discrete observations using the phase-jump technique described by \cite{dracula}. 

Following the development of these initial timing solutions, each observation was then iteratively re-fit in order to generate progressively more precise TOAs, which in turn allowed for the production of more precise timing solutions. Each new set of TOAs was partially summed alternatingly in either the time or frequency domains in order to also fit for each pulsar's dispersion measure (DM). Once no further significant improvements in precision could be made, the refined timing solutions (as listed in Tables~\ref{tab: timing table 1}, \ref{tab: timing table 2} and \ref{tab: timing table 3}) were then produced using \textsc{tempo2} and a final set of time-domain TOAs. 

\begin{table*}
\begin{center}
\caption{Best-fit \textsc{tempo2} timing parameters for four FAST-discovered pulsars, PSRs~J0021$-$0909, J0344$-$0901, J0529$-$0715 and J0803$-$0942. Each model uses a DE421 planetary ephemeris and TCB (Barycentric Coordinate Time) time units. Values in parentheses represent 1-$\sigma$ uncertainties on the final digit or digits after weighting the TOAs using the provided EFAC values such that the reduced $\chi^{2}=1$. DM distances are calculated according to the NE2001 model \citep{NE2001a} and the YMW16 model \citep{YMW16}. $\tau_\text{c}$, $B_\text{surf}$ and $\dot{E}$ are calculated according to the equations presented in \citet{lk05}.}\label{tab: timing table 1}
\begin{tabular}{lllll}
\hline
Pulsar name & PSR~J0021$-$0909 & PSR~J0344$-$0901 & PSR~J0529$-$0715 & PSR~J0803$-$0942\\
\hline
\textit{Measured parameters} \\
Right ascension, $\alpha$ (J2000)\dotfill & 00:21:51.47(3) & 03:44:37.471(2) & 05:29:08.973(2) & 08:03:26.848(13) \\
Declination, $\delta$ (J2000)\dotfill & $-$09:09:58.7(11) & $-$09:01:02.66(10) & $-$07:15:26.43(10) & $-$09:42:50.81(14) \\
Spin frequency, $\nu$ ($\text{s}^{-1}$)\dotfill & 0.43212768588(3) & 0.815656977709(8) & 1.450907946316(17) & 1.75052694621(4) \\
Spin frequency derivative, $\dot{\nu}$ ($\text{s}^{-2}$)\dotfill & $-1.94(3)\times10^{-16}$ & $-2.3091(10)\times10^{-15}$ & $-1.2114(17)\times10^{-15}$ & $-3.558(5)\times10^{-15}$ \\
Dispersion measure, DM ($\text{cm}^{-3}\,\text{pc}$)\dotfill & 25.2(10) & 30.9(3) & 87.3(4) & 21.1(3) \\
\\
\textit{Fitting parameters} \\
First TOA (MJD)\dotfill & 58194.1 & 58199.2 & 58199.2 & 58314.2 \\
Last TOA (MJD)\dotfill & 58715.8 & 58715.9 & 58710.9 & 58705.2 \\
Timing epoch (MJD)\dotfill & 58388.0 & 58393.0 & 58372.0 & 58410.0 \\
Number of TOAs\dotfill & 102 & 586 & 203 & 199 \\
Total integration time (hr) & 20.8 & 61.1 & 54.2 & 61.1 \\
Weighted RMS residual ($\mu\text{s}$)\dotfill & 1390 & 1264 & 790 & 466 \\
TOA weighting factor, EFAC\dotfill & 1.82 & 1.54 & 1.60 & 1.04 \\
\\
\textit{Derived parameters} \\
Galactic longitude, $l$ ($^\circ$)\dotfill & 100.290 & 197.397 & 210.065 & 230.070 \\
Galactic latitude, $b$ ($^\circ$)\dotfill & $-$70.726 & $-$45.333 & $-$21.582 & 11.197 \\
Spin period, $P$ (ms)\dotfill & 2314.13082912(16) & 1226.005572598(12) & 689.223601359(8) & 571.256559156(14) \\
Spin period derivative, $\dot{P}$\dotfill & $1.039(16)\times10^{-15}$ & $3.4708(15)\times10^{-15}$ & $5.755(8)\times10^{-16}$ & $1.1612(16)\times10^{-15}$ \\
DM distance, $d$ (kpc) \\
\multicolumn{1}{r}{NE2001} & 1.3 & 1.5 & $>46$ & 1.3 \\
\multicolumn{1}{r}{YMW16} & $>25$ & 1.5 & 7.0 & 0.8 \\
Characteristic age, $\tau_\text{c}$ (Myr)\dotfill & 35.2 & 5.58 & 18.9 & 7.77 \\
Surface magnetic field, $B_\text{surf}$ ($10^{10}\,\text{G}$)\dotfill & 155 & 206 & 63.0 & 81.4 \\
Spin-down luminosity, $\dot{E}$ ($10^{30}\,\text{erg}\,\text{s}^{-1}$)\dotfill & 3.31 & 74.4 & 69.4 & 246 \\
\hline
\end{tabular}
\end{center}
\end{table*}

\begin{table*}
\begin{center}
\caption{Best-fit \textsc{tempo2} timing parameters for four FAST-discovered pulsars, PSRs~J1851$-$0633, J1900$-$0134, J1919$+$2621 and J1926$-$0652. Details as per Table~\ref{tab: timing table 1}.}\label{tab: timing table 2}
\begin{tabular}{lllll}
\hline
Pulsar name & PSR~J1851$-$0633$^\text{a}$ & PSR~J1900$-$0134 & PSR~J1919$+$2621 & PSR~J1926$-$0652 \\
\hline
\textit{Measured parameters} \\
Right ascension, $\alpha$ (J2000)\dotfill & 18:51:41.646(5) & 19:00:26.174(5) & 19:19:41.013(4) & 19:26:37.041(6) \\
Declination, $\delta$ (J2000)\dotfill & $-$06:33:48.4(2) & $-$01:34:38.2(3) & $+$26:21:29.46(9) & $-$06:52:42.7(4) \\
Spin frequency, $\nu$ ($\text{s}^{-1}$)\dotfill & 0.520751388080(6) & 0.545752603980(8) & 1.53489092390(5) & 0.621575004134(7) \\
Spin frequency derivative, $\dot{\nu}$ ($\text{s}^{-2}$)\dotfill & $-6.502(13)\times10^{-16}$ & $-9.0812(13)\times10^{-15}$ & $-1.286(5)\times10^{-15}$ & $-1.512(7)\times10^{-16}$ \\
Dispersion measure, DM ($\text{cm}^{-3}\,\text{pc}$)\dotfill & 229.1(7) & 178.1(5) & 96.5(2) & 85.3(7) \\
\\
\textit{Fitting parameters} \\
First TOA (MJD)\dotfill & 58221.7 & 58007.4 & 58271.7 & 58034.2 \\
Last TOA (MJD)\dotfill & 58633.7 & 58531.9 & 58632.7 & 58695.7 \\
Timing epoch (MJD)\dotfill & 58382.0 & 58269.0 & 58333.0 & 58308.0 \\
Number of TOAs\dotfill & 63 & 117 & 274 & 463 \\
Total integration time (hr) & 15.0 & 21.6 & 6.8 & 46.4 \\
Weighted RMS residual ($\mu\text{s}$)\dotfill & 469 & 827 & 550 & 2652 \\
TOA weighting factor, EFAC\dotfill & 0.92 & 1.10 & 1.22 & 1.20 \\
\\
\textit{Derived parameters} \\
Galactic longitude, $l$ ($^\circ$)\dotfill & 27.114 & 32.553 & 59.535 & 30.751 \\
Galactic latitude, $b$ ($^\circ$)\dotfill & $-$3.044 & $-$2.721 & 5.993 & $-$10.936 \\
Spin period, $P$ (ms)\dotfill & 1920.30213052(2) & 1832.33207264(3) & 651.51209407(2) & 1608.816302697(18) \\
Spin period derivative, $\dot{P}$ \dotfill & $2.398(5)\times10^{-15}$ & $3.0490(4)\times10^{-14}$ & $5.46(2)\times10^{-16}$ & $3.914(17)\times10^{-16}$ \\
DM distance, $d$ (kpc) \\
\multicolumn{1}{r}{NE2001} & 5.0 & 4.5 & 4.5 & 2.9 \\
\multicolumn{1}{r}{YMW16} & 6.2 & 4.9 & 6.1 & 5.3 \\
Characteristic age, $\tau_\text{c}$ (Myr)\dotfill & 12.7 & 0.950 & 18.9 & 65.0 \\
Surface magnetic field, $B_\text{surf}$ ($10^{10}\,\text{G}$)\dotfill & 215 & 747 & 59.6 & 79.4 \\
Spin-down luminosity, $\dot{E}$ ($10^{30}\,\text{erg}\,\text{s}^{-1}$)\dotfill & 13.4 & 196 & 78.0 & 3.71 \\
\hline
\multicolumn{5}{l}{$^\text{a}$ As noted previously, PSR~J1851$-$0633 was published in \cite{ccb+20} as PSR~J1851$-$06, but without a timing solution.} \\
\end{tabular}
\end{center}
\end{table*}

\begin{table*}
\begin{center}
\caption{Best-fit \textsc{tempo2} timing parameters for three FAST-discovered pulsars, PSRs~J1931$-$0144, J1945$+$1211 and J2323$+$1214. Details as per Table~\ref{tab: timing table 1}.}\label{tab: timing table 3}
\begin{tabular}{lllll}
\hline
Pulsar name & PSR~J1931$-$0144 & PSR~J1945$+$1211 & PSR~J2323$+$1214 \\
\hline
\textit{Measured parameters} \\
Right ascension, $\alpha$ (J2000)\dotfill & 19:31:32.025(9) & 19:45:11.56(3) & 23:23:21.619(7) \\
Declination, $\delta$ (J2000)\dotfill & $-$01:44:22.5(4) & $+$12:11:46.2(7) & $+$12:14:12.70(16) \\
Spin frequency, $\nu$ ($\text{s}^{-1}$)\dotfill & 1.68446199761(3) & 0.21022552160(2) & 0.265993420483(5) \\
Spin frequency derivative, $\dot{\nu}$ ($\text{s}^{-2}$)\dotfill & $-8.3(3)\times10^{-17}$ & $-1.67(2)\times10^{-16}$ & $-1.307(4)\times10^{-16}$ \\
Dispersion measure, DM ($\text{cm}^{-3}\,\text{pc}$)\dotfill & 38.3(13) & 92.7(16) & 29.0(3) \\
\\
\textit{Fitting parameters} \\
First TOA (MJD)\dotfill & 58034.2 & 58051.2 & 58151.2\\
Last TOA (MJD)\dotfill & 58655.8 & 58642.7 & 58707.7\\
Timing epoch (MJD)\dotfill & 58282.0 & 59298.0 & 58315.0\\
Number of TOAs\dotfill & 87 & 144 & 98 \\
Total integration time (hr) & 15.9 & 8.8 & 20.9 \\
Weighted RMS residual ($\mu\text{s}$)\dotfill & 1755 & 5185 & 508 \\
TOA weighting factor, EFAC\dotfill & 1.21 & 2.85 & 1.43 \\
\\
\textit{Derived parameters} \\
Galactic longitude, $l$ ($^\circ$)\dotfill & 35.974 & 49.984 & 91.595 \\
Galactic latitude, $b$ ($^\circ$)\dotfill & $-$9.711 & $-$6.076 & $-$45.209 \\
Spin period, $P$ (ms)\dotfill & 593.661359782(12) & 4756.7963794(5) & 3759.49148736(8)\\
Spin period derivative, $\dot{P}$ \dotfill & $2.93(12)\times10^{-17}$ & $3.79(5)\times10^{-15}$ & $1.847(5)\times10^{-15}$ \\
DM distance, $d$ (kpc) \\
\multicolumn{1}{r}{NE2001} & 1.7 & 3.9 & 1.7 \\
\multicolumn{1}{r}{YMW16} & 1.4 & 3.8 & $>25$ \\
Characteristic age, $\tau_\text{c}$ (Myr)\dotfill & 320 & 19.8 & 32.2 \\
Surface magnetic field, $B_\text{surf}$ ($10^{10}\,\text{G}$)\dotfill & 13.2 & 425 & 264 \\
Spin-down luminosity, $\dot{E}$ ($10^{30}\,\text{erg}\,\text{s}^{-1}$)\dotfill & 5.5 & 1.40 & 1.37 \\
\hline
\end{tabular}
\end{center}
\end{table*}

\begin{figure}
    \begin{center}
    \includegraphics[width=\columnwidth]{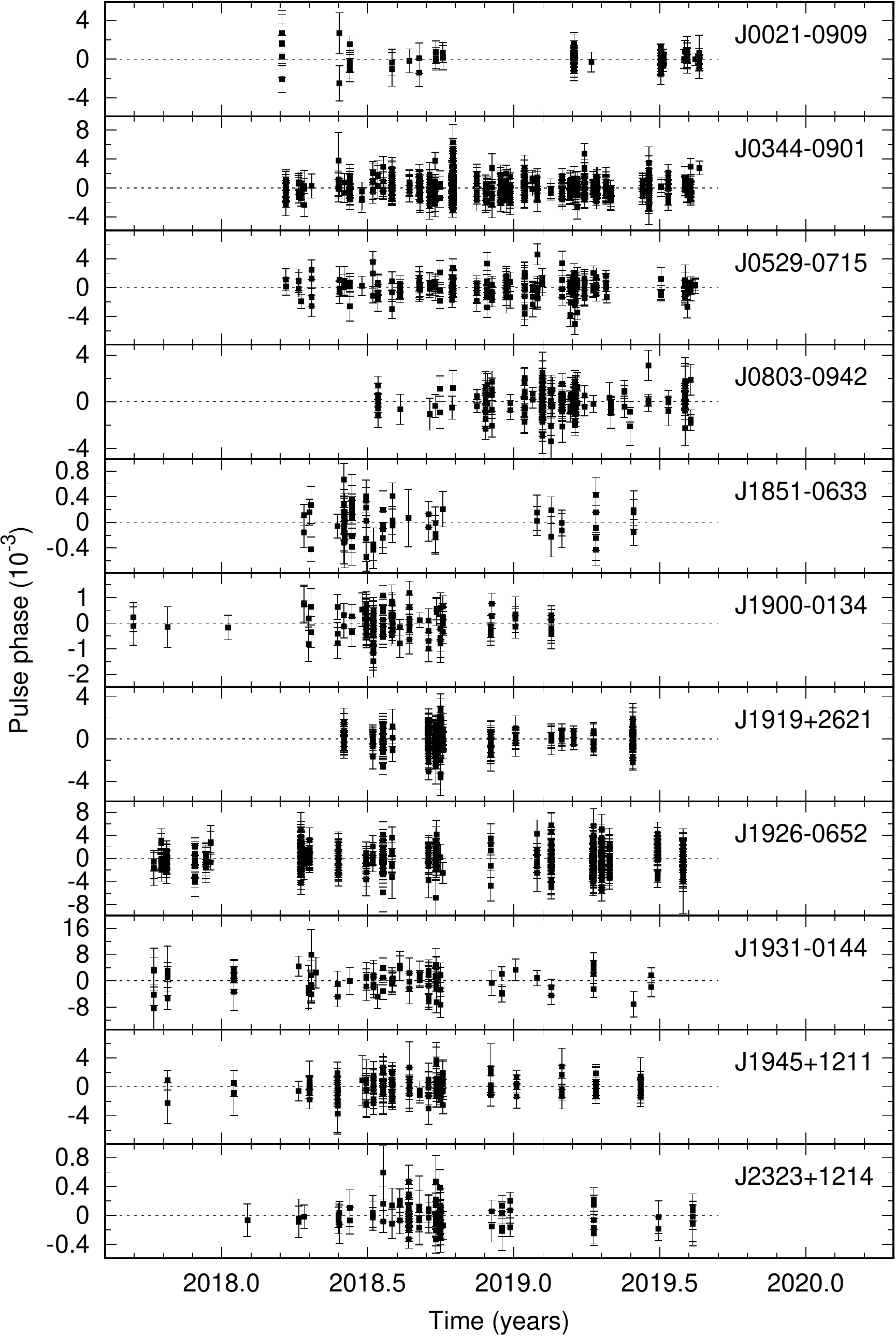}
    \caption{\textsc{tempo2} timing residuals for \PSRnumPKSFAST pulsars, based upon the timing solutions listed in Tables~\ref{tab: timing table 1}, \ref{tab: timing table 2} and \ref{tab: timing table 3}. In each case, the TOA error bars have been weighted such that the reduced $\chi^2=1$. The dashed line in each plot represents the zero-line, and corresponds to the arrival times predicted by each timing solution. Note that the vertical scale (given in units of pulse phase) differs for each pulsar.}\label{fig: residuals}
    \end{center}
\end{figure}

The residuals from these timing models are shown in Figure~\ref{fig: residuals}, after having been weighted such that their reduced $\chi^{2}=1$ (Tables~\ref{tab: timing table 1}, \ref{tab: timing table 2} and \ref{tab: timing table 3} provide the \textsc{tempo2} EFAC weighting values). Each set of TOAs appears to be distributed in an approximately Gaussian manner, with no remaining systematic trends, indicating that each model is well-fit. There is also no apparent long-term timing noise in any of the \PSRnumPKSFAST pulsars. This remains true even when each observation is summed fully in time in order to reduce the number of TOAs and increase the available timing precision. However, given the fact that no pulsar in this set has been observed continuously for longer than $\sim2\,\text{yr}$, such timing noise is likely to become detectable with longer observing timescales.

Estimates of the distance to each pulsar based upon their measured DM values were calculated according to both the NE2001 model \citep{NE2001a} and the YMW16 model \citep{YMW16} of the Galactic free electron density, and are also provided in Tables~\ref{tab: timing table 1}, \ref{tab: timing table 2} and \ref{tab: timing table 3}. We do not attempt to evaluate the relative accuracy of these models in this paper, and present both models as a reflection of the large uncertainties typically involved in DM-distance estimation. This fact is highlighted in the case of PSRs~J0021$-$0909, J0529$-$0715 and J2323$+$1214, for which at least one model indicates a DM-distance to the pulsar exceeding the model's limits along that particular line of sight, which if true would place the pulsar outside of our Galaxy. As we have no additional evidence to believe that any of these pulsars is extragalactic in nature, we instead attribute these particular distance estimates to inaccuracies in each model.

\begin{figure*}
\begin{center}
\includegraphics[height=0.49\textwidth, angle=270]{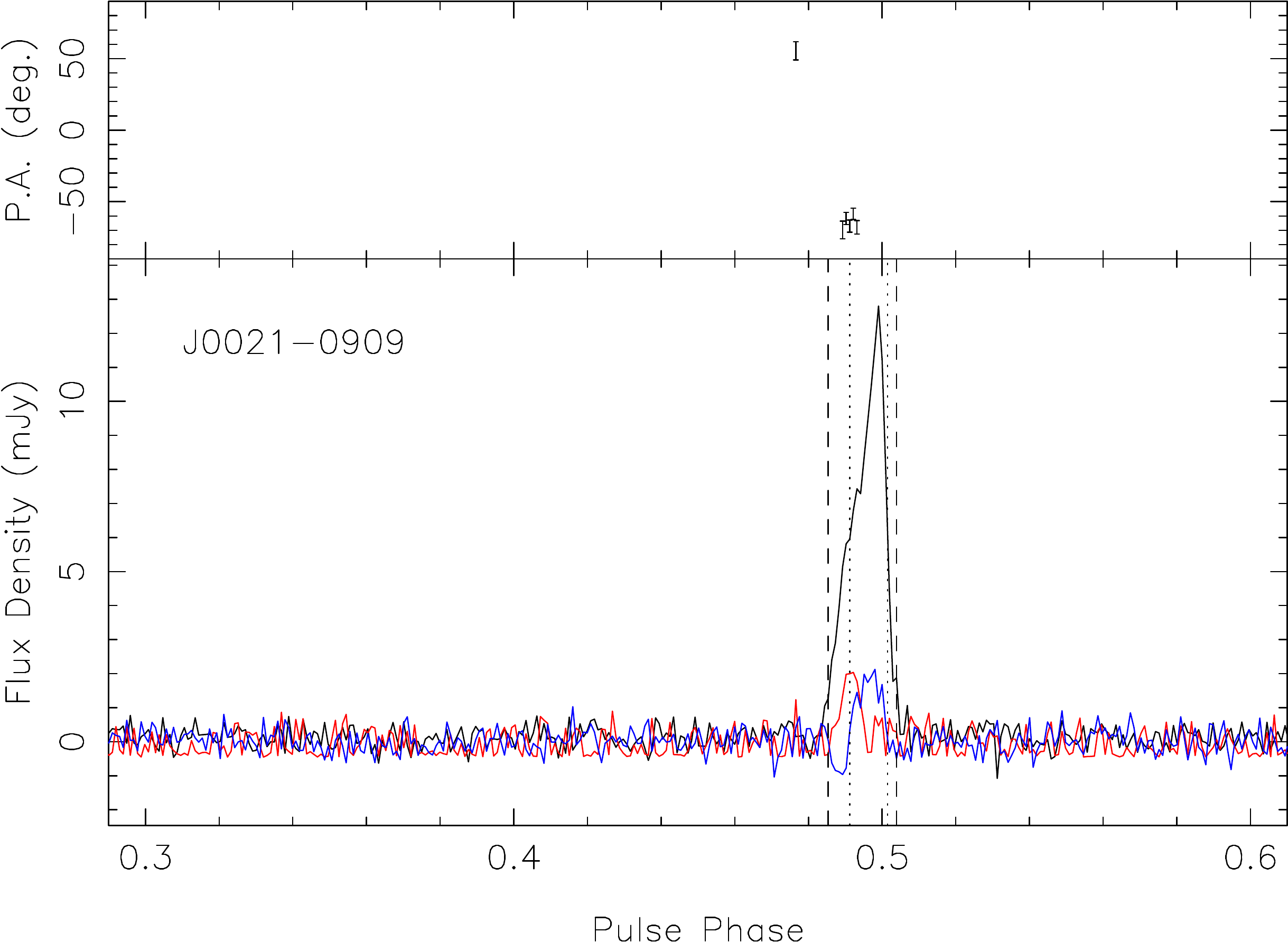}
\vspace{0.5cm}
\includegraphics[height=0.49\textwidth, angle=270]{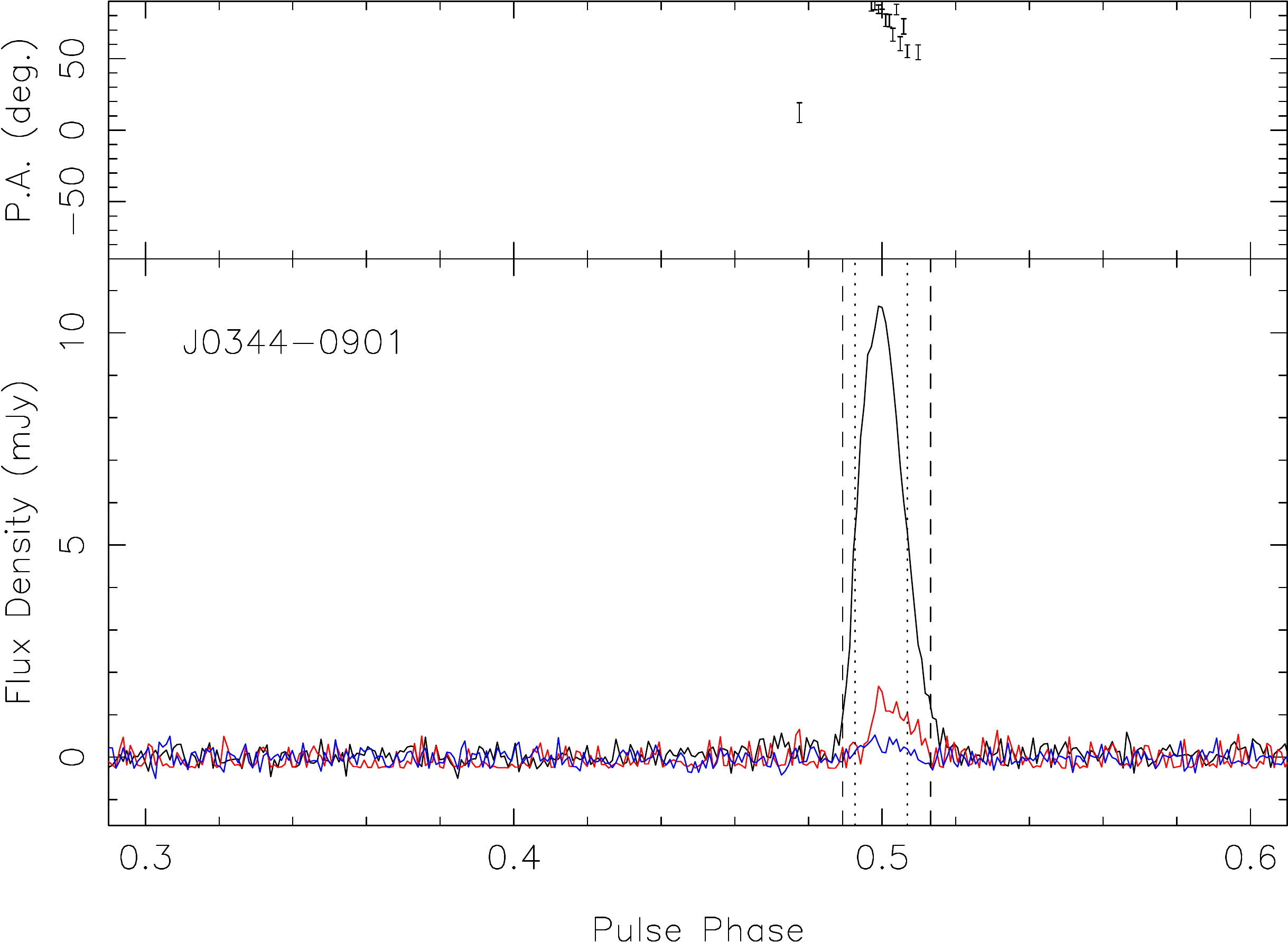}
\includegraphics[height=0.49\textwidth, angle=270]{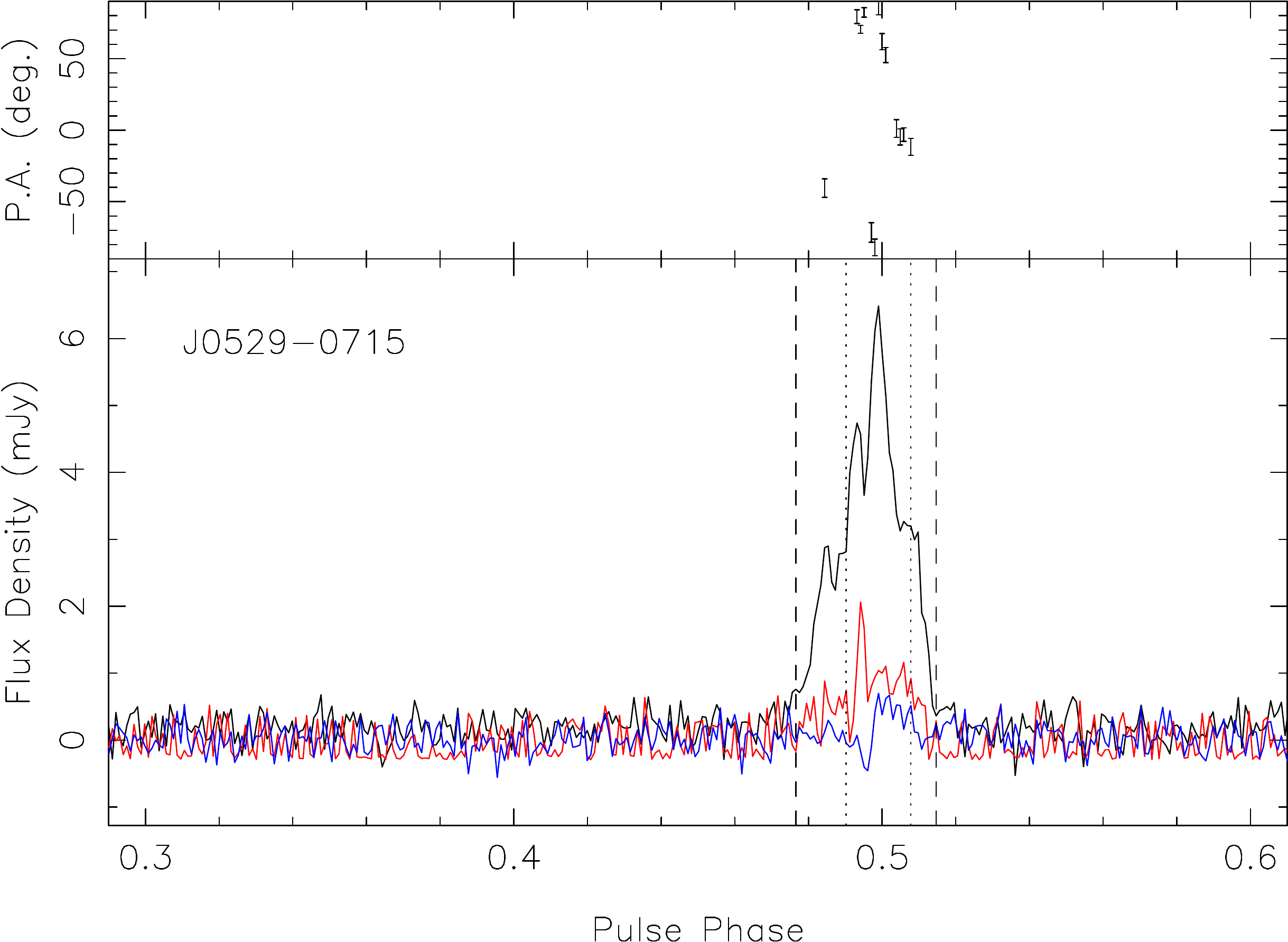}
\vspace{0.5cm}
\includegraphics[height=0.49\textwidth, angle=270]{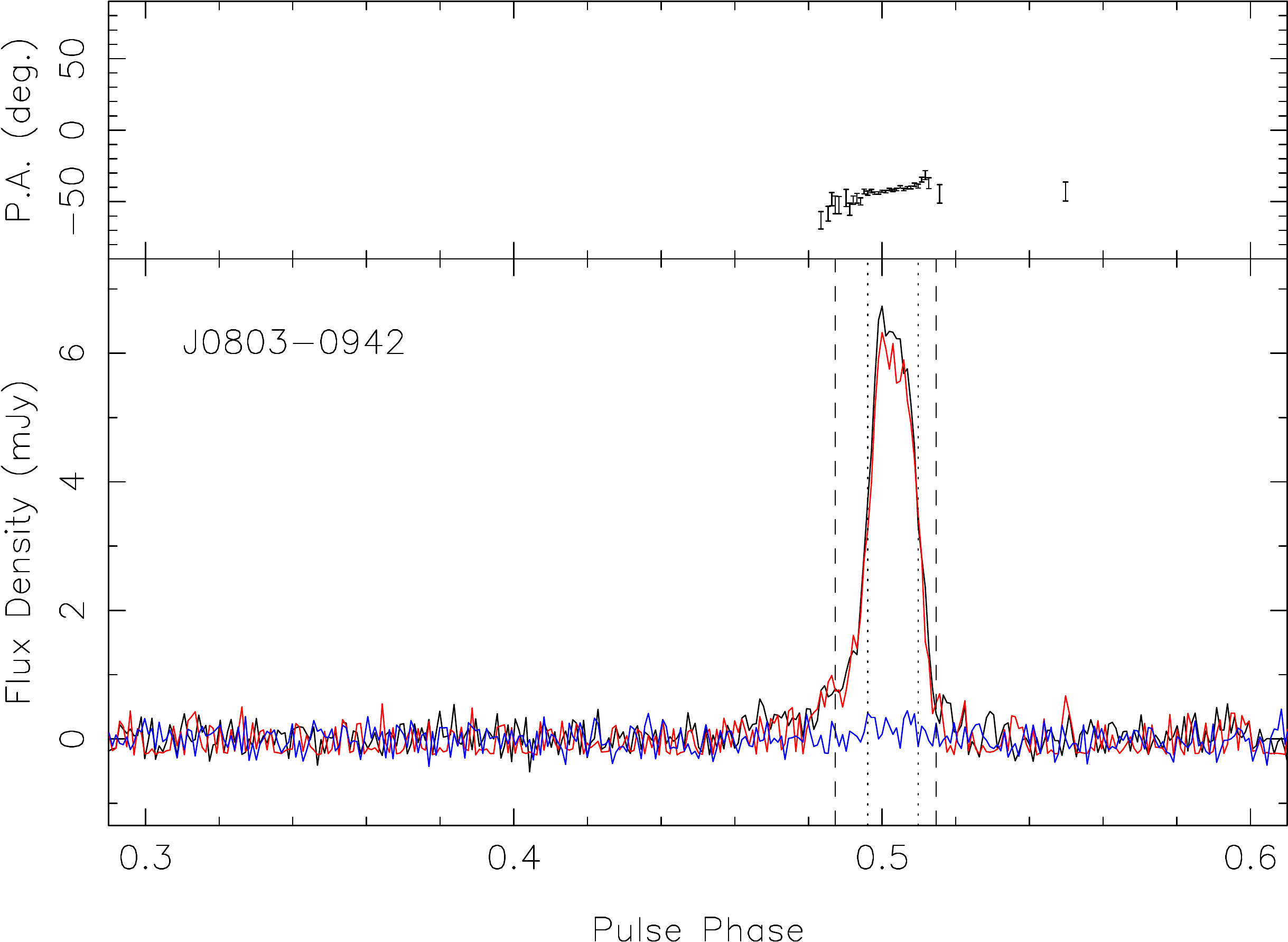}
\includegraphics[height=0.49\textwidth, angle=270]{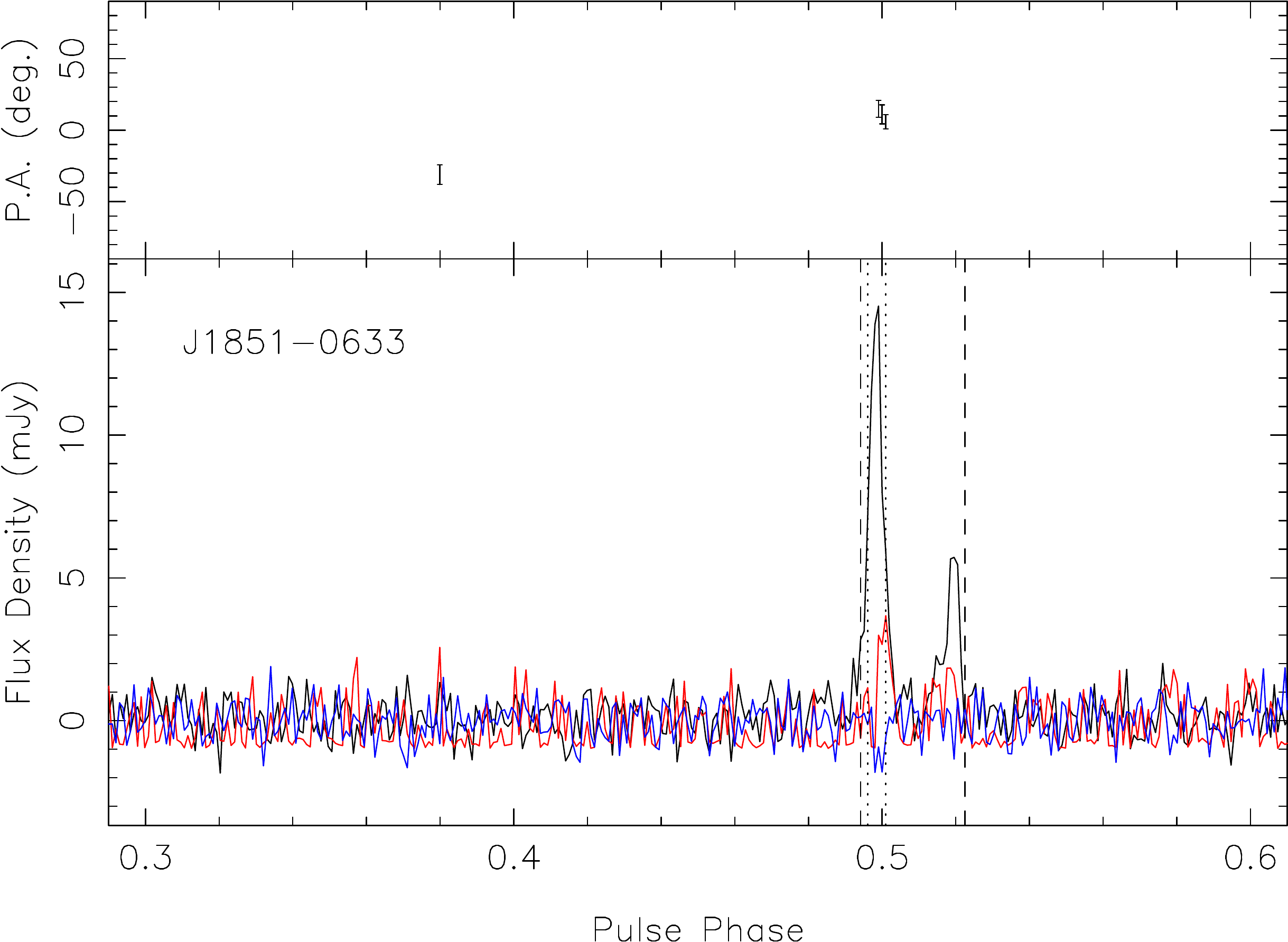}
\vspace{0.5cm}
\includegraphics[height=0.49\textwidth, angle=270]{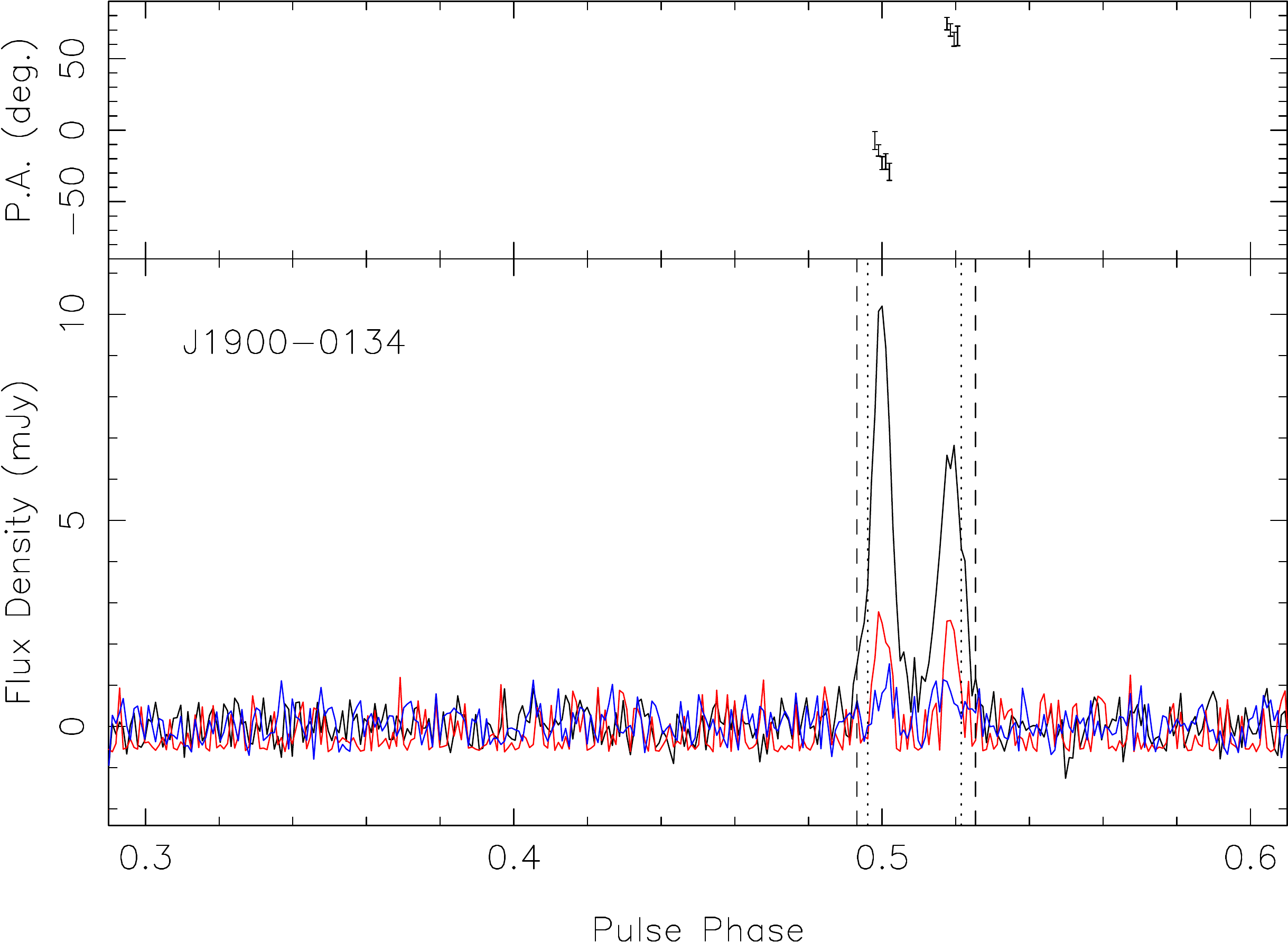}
\end{center}
\caption{Integrated pulse profiles for six FAST-discovered pulsars, PSRs~J0021$-$0909, J0344$-$0901, J0529$-$0715, J0803$-$0942, J1851$-$0633 and J1900$-$0134. The total integration time used to construct each profile is provided in Table~\ref{tab: polarisation table}. Each profile has been rotated so that its peak is at a pulse-phase of 0.5. Total intensity ($I$) is shown in black, while the linearly-polarised ($L$) and circularly-polarised ($V$) components are shown in red and blue respectively. The approximate widths of each profile's width (as per Table~\ref{tab: polarisation table}) are marked, including the $W_{10}$ (large vertical dash) and $W_{50}$ (small vertical dash). Above each plot is shown the changing polarisation angle (PA) of the linearly-polarised flux, referenced to a central frequency of 1369\,MHz.}\label{fig: pulse profiles 1}
\end{figure*}

\begin{figure*}
\begin{center}
\includegraphics[height=0.49\textwidth, angle=270]{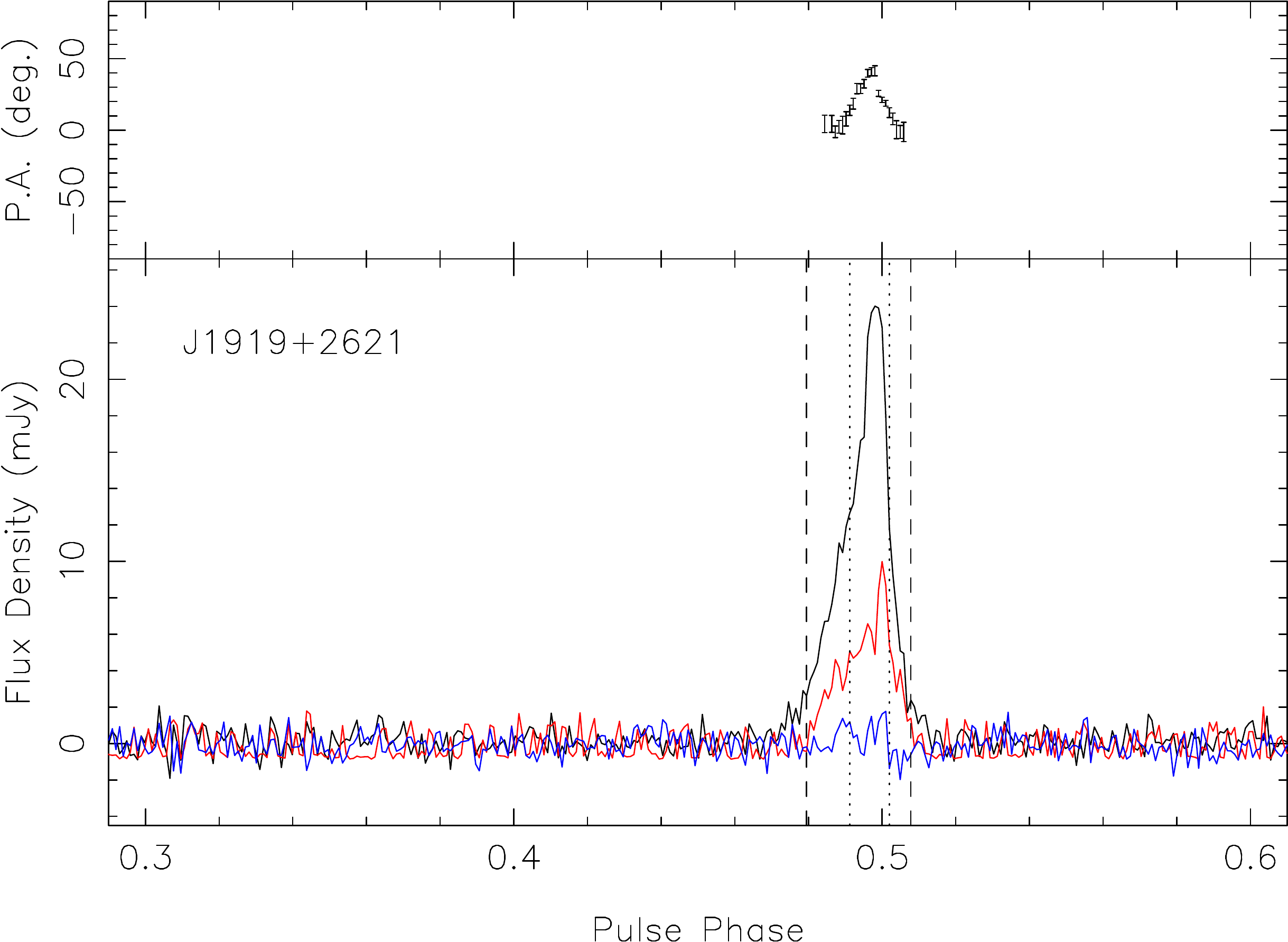}
\vspace{0.5cm}
\includegraphics[height=0.49\textwidth, angle=270]{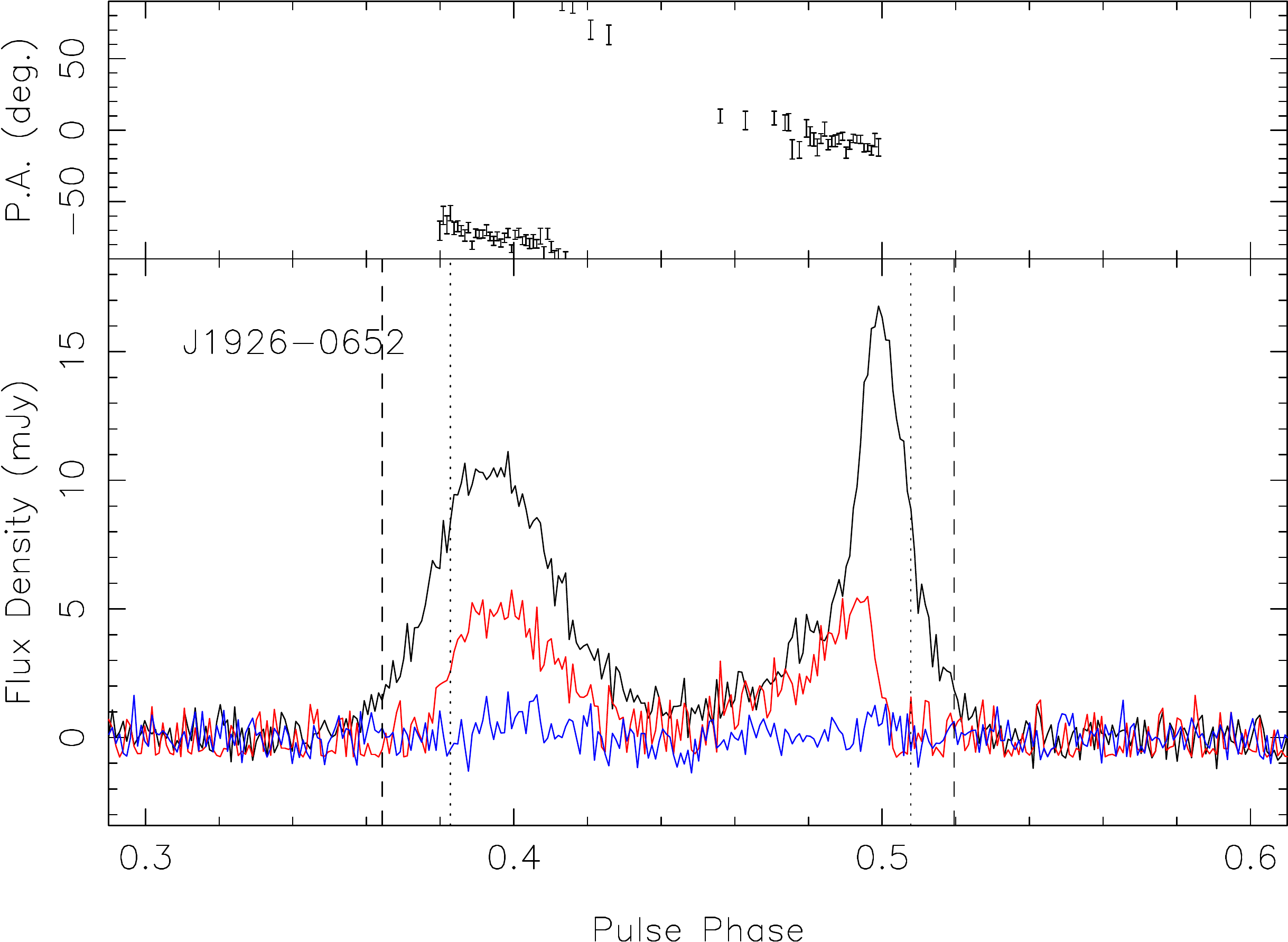}
\includegraphics[height=0.49\textwidth, angle=270]{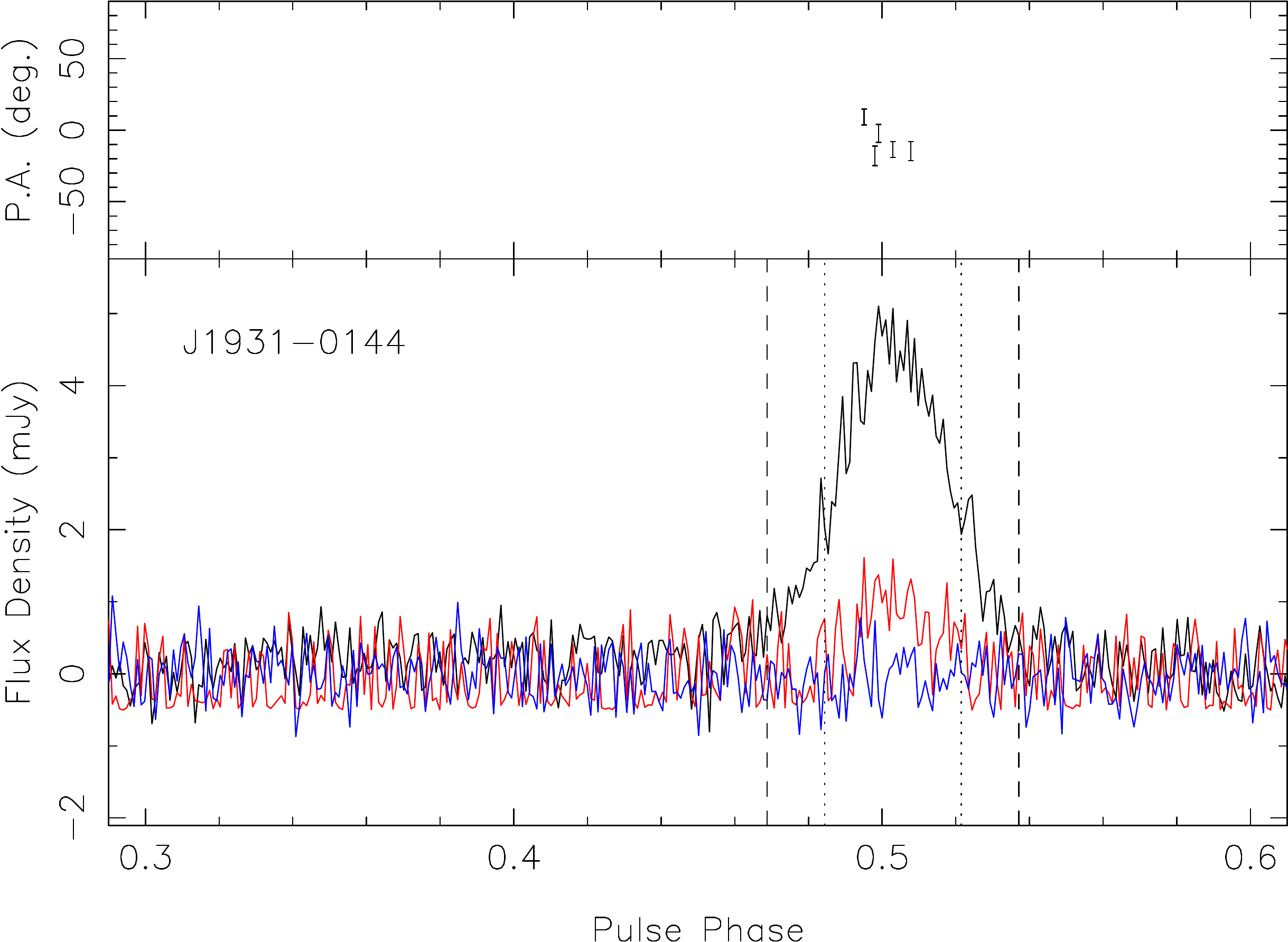}
\vspace{0.5cm}
\includegraphics[height=0.49\textwidth, angle=270]{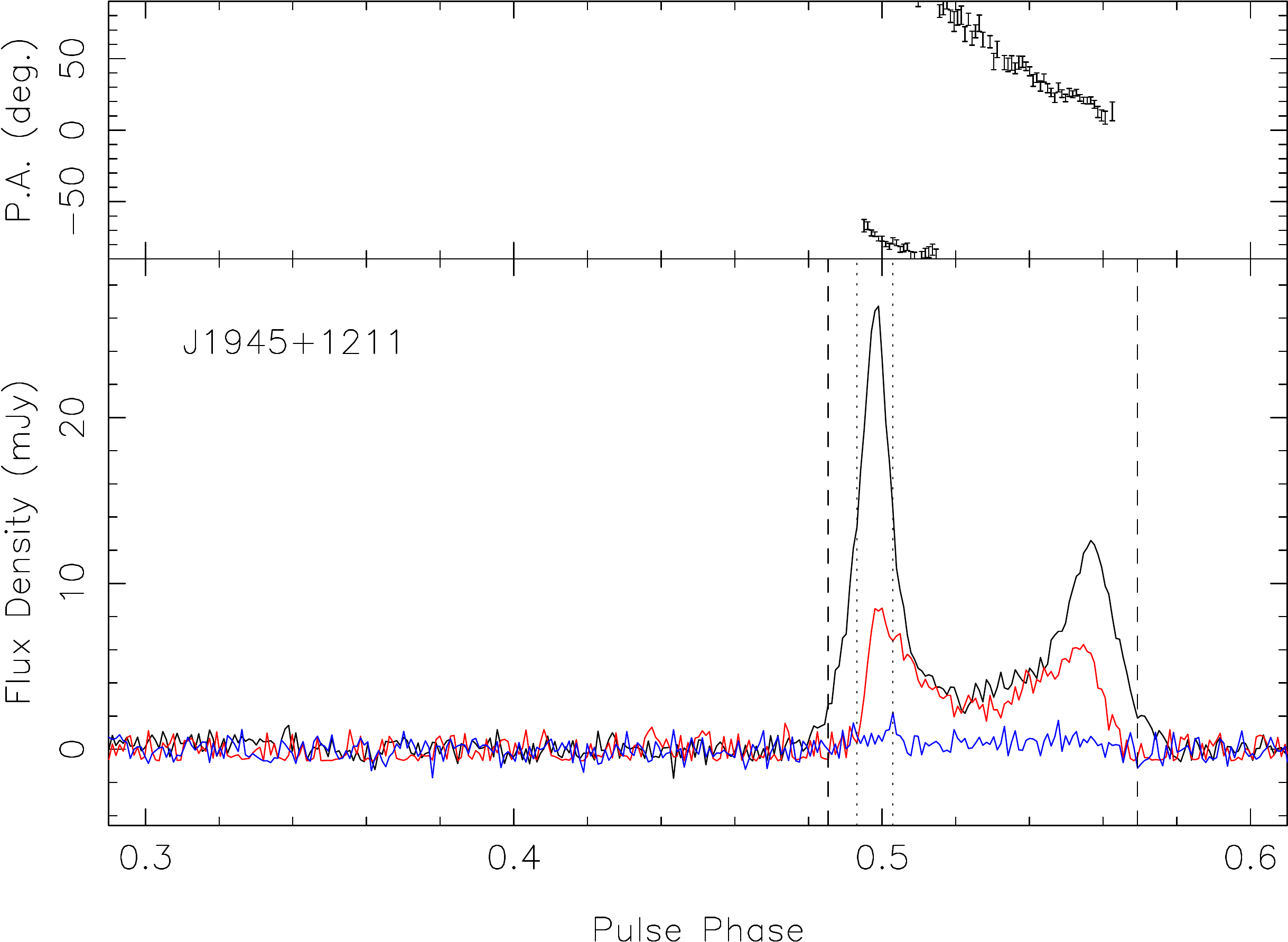}
\includegraphics[height=0.49\textwidth, angle=270]{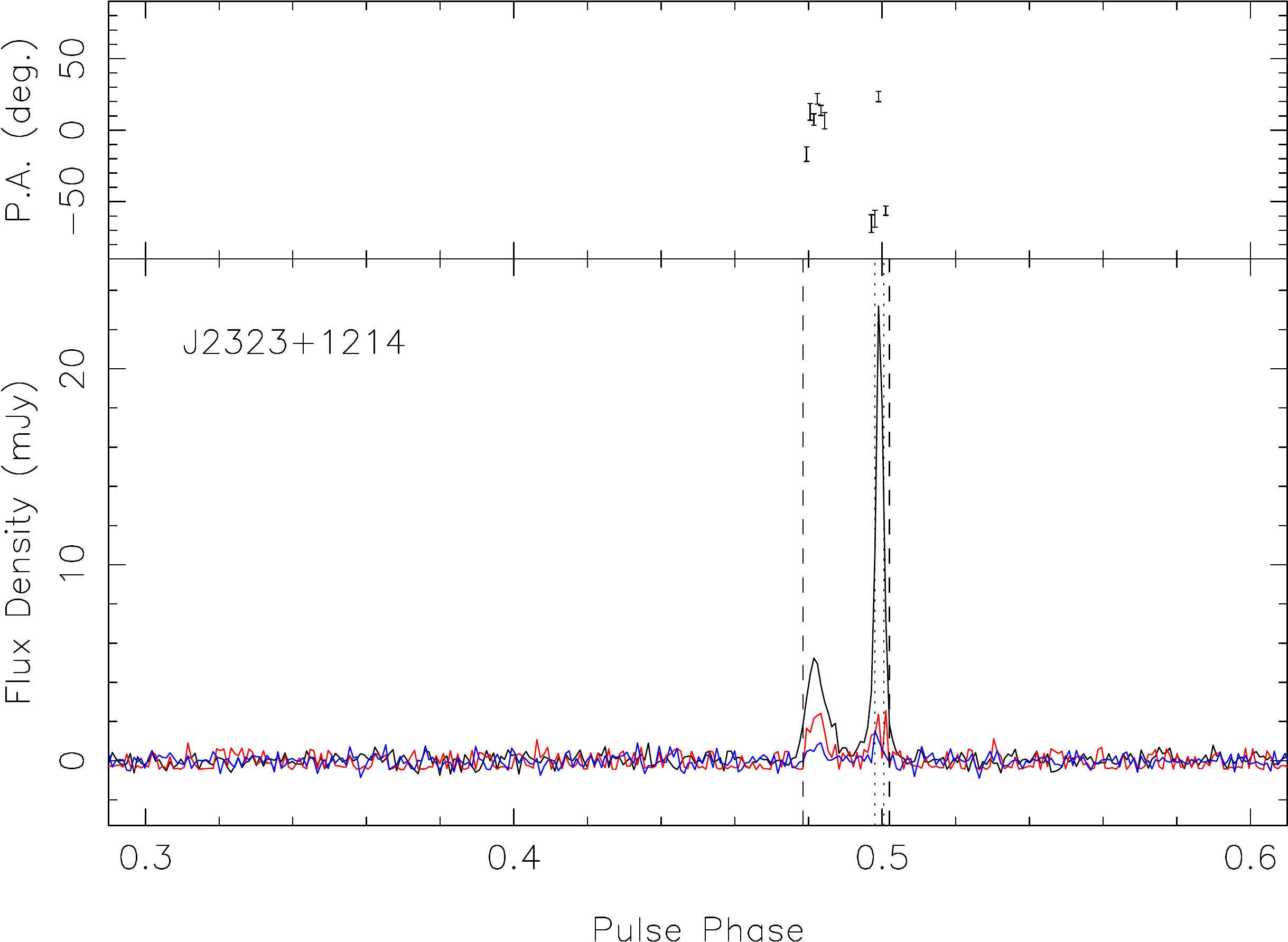}
\vspace{0.5cm}
\end{center}
\caption{Integrated pulse profiles for five FAST-discovered pulsars, PSRs~J1919$+$2621, J1926$-$0652, J1931$-$0144, J1945$+$1211 and J2323$+$1214. Details as per Figure~\ref{fig: pulse profiles 1}.}\label{fig: pulse profiles 2}
\end{figure*}

Integrated pulse profiles for each of the \PSRnumPKSFAST pulsars are shown in Figures~\ref{fig: pulse profiles 1} and \ref{fig: pulse profiles 2}. These profiles were constructed by summing each pulsar's calibrated observations in time on a per-receiver basis\footnote{Differences in the header information between the files produced by the MB20 and UWL receivers prevented them being easily summed together.}, using each pulsar's timing solution to ensure coherent phase-alignment. The resulting summed profile with the highest S/N was then selected for further analysis. The total integration time of the observations used to construct this chosen file can be found in Table~\ref{tab: polarisation table}. The rotation measure (RM) of each pulsar was determined and applied using the \textsc{rmfit} software package before further summing each file in frequency. The resulting time- and frequency-summed profiles were then plotted using \textsc{pav} after having been rotated such that the profile peak was at a pulse phase of approximately 0.5.  

Further details regarding the pulsed emission of each pulsar are listed in Table~\ref{tab: polarisation table}. Pulse widths at 50\,\% and 10\,\% of the pulse peak ($W_{50}$ and $W_{10}$ respectively) were measured by modeling each profile in \textsc{paas} using a series of Gaussian curves, against which the pulse widths were measured using \textsc{pdv}. The flux density at 1400\,MHz ($S_{1400}$) was calculated using the \textsc{jflux} software package, which allows a user-defined on-pulse region. Prior to the derivation of a timing solution, each pulsar's position was only known to within a precision of approximately $0.12^\circ$. Consequentially, each pulsar was typically observed at an offset from its true position, which would cause a reduction in its apparent flux density. We corrected for this offset following the method of \cite{ncb15}, wherein we modelled the primary beam within the FWHM as a Gaussian and corrected the measured flux value $S_\text{obs}$ according to the expression
\begin{equation}\label{eqn: flux correction}
    S_{1400} = S_\text{obs}\text{exp}\left(\frac{\theta^2}{2\sigma^2}\right),
\end{equation} where $\theta$ is the angular offset between the position of the pulsar and the telescope, and $\sigma$ is given by
\begin{equation}\label{eqn: sigma}
    \sigma=\frac{\text{FWHM}}{2\sqrt{2\ln{2}}}.
\end{equation} For both the MB20 and UWL receivers, $\text{FWHM}\simeq0.24^\circ$ at 1400\,MHz, therefore $\sigma\simeq0.1^\circ$.

\begin{table*}
\begin{center}
\caption{Pulse profile, flux density and polarisation properties of \PSRnumPKSFAST FAST-discovered pulsars. Listed for each pulsar are its flux density and luminosity at 1400\,MHz ($S_{1400}$ and $L_{1400}$ respectively), pulse width at 50\,\% and 10\,\% of the profile peak ($W_{50}$ and $W_{10}$ respectively), rotation measure (RM), and fractional linear, circular and absolute circular polarisation ($L/I$, $V/I$ and $\left|V\right|/I$ respectively). Luminosities are calculated according to the NE2001 model \citep[left column;][]{NE2001a} and the YMW16 model \citep[right column;][]{YMW16}. Also listed are the receivers and total integration time ($\sum t_\text{obs}$) used to construct the integrated pulse profiles shown in Figures~\ref{fig: pulse profiles 1} and \ref{fig: pulse profiles 2}. Values in parentheses represent 1-$\sigma$ uncertainties on the final digit or digits.}\label{tab: polarisation table}
\begin{tabular}{llcccccccccc}
\hline
PSR & $S_{1400}$ & \multicolumn{2}{c}{$L_{1400}$} &$W_{50}$ & $W_{10}$  & RM & $L/I$ & $V/I$ & $\left|V\right|/I$ & Rcvr. & $\sum t_\text{obs}$ \\
 & (mJy) & \multicolumn{2}{c}{($\text{mJy}\,\text{kpc}^{2}$)} & (ms) & (ms) & ($\text{rad}\,\text{m}^{-2}$) & (per cent) & (per cent) & (per cent) & & (h) \\
\hline
J0021$-$0909 & 0.134(11) & 0.21 & $>84$ & 23.8 & 42.7 & $-$15(18) & 14.6(18) & 7.7(12) & 11.8(7) & UWL & 15.0 \\
J0344$-$0901 & 0.155(7) & 0.36 & 0.35 & 17.1 & 24.5 & 7(11) & 12.3(9) & 2.7(6) & 1.3(4) & UWL & 41.7 \\
J0529$-$0715 & 0.125(8) & $>260$ & 6.1 & 11.9 & 26.3 & $-$46(3) & 25(2) & 4.4(12) & 1.9(7) & UWL & 32.1 \\
J0803$-$0942 & 0.102(7) & 0.18 & 0.068 & 8.21 & 15.8 & $-$17.5(5) & 95.5(18) & 4.1(10) & 1.3(6) & UWL & 48.4 \\
J1851$-$0633 & 0.12(2) & 2.9 & 4.6 & 9.16 & 55.0 & 513(8) & 33(6) & $-$4(4) & 1(2) & MB20 & 12.0 \\
J1900$-$0134 & 0.202(15) & 4.0 & 4.8 & 45.3 & 57.5 & 20(13) & 25(3) & 10.0(19) & 5.3(11) & MB20 & 19.3 \\
J1919$+$2621 & 0.57(3) & 12 & 21 & 7.11 & 18.6 & 63.4(12) & 36.0(16) & 1.4(1.0) & 2.3(6) & MB20 & 3.7 \\
J1926$-$0652 & 1.01(3)$^{\text{a}}$ & 8.6 & 28 & 202 & 251 & $-$54(3) & 48.1(16) & 1.7(9) & 1.2(6) & MB20 & 28.3 \\
J1931$-$0144 & 0.196(14) & 0.57 & 0.38 & 22.2 & 40.5 & $-$50(3) & 28(3) & $-$1.4(19) & $-$0.6(11) & MB20 & 11.3 \\
J1945$+$1211 & 0.634(19) & 9.7 & 9.0 & 48.8 & 396 & $-$21.4(7) & 51.6(13) & 5.8(8) & 2.0(5) & MB20 & 7.0 \\
J2323$+$1214 & 0.144(18) & 0.40 & $>90$ & 9.62 & 86.9 & $-$17(6) & 24(2) & 8.2(15) & 2.6(9) & MB20 & 16.9 \\
\hline
\multicolumn{12}{l}{\footnotesize{$^\text{a}$ Measured using only the portions of each observation when the pulsar's emission was seen to be `on'. See \cite{zlh+19} for}}\\
\multicolumn{12}{l}{\footnotesize{further details.}}\\
\end{tabular}
\end{center}
\end{table*}

\subsection{Polarisation}\label{subsec: polarisation}

In order to measure the polarisation properties of each profile, we used the \textsc{psrsalsa}\footnote{https://github.com/weltevrede/psrsalsa} software package \citep{psrsalsa}. The \textsc{ppol} tool was used to extract the Stokes $I_i$ (total intensity) and $V_i$ (circular polarisation) values of each profile bin, as well as the bias-corrected value of $L_i$ (linear polarisation). For each parameter, the off-pulse root-mean-squared (RMS) was also calculated, here termed as $\sigma_I$, $\sigma_L$ and $\sigma_V$ respectively\footnote{As bias-correction involves zeroing the mean of each parameter, the RMS measurement is equivalent to the standard deviation $\sigma$.}. The $W_{10}$ of each profile was then used to determine a precise on-pulse region, over which the values of $I_i$, $L_i$ and $V_i$ were then summed, such that
\begin{equation}\label{eqn: sumI}
    I = \sum^{N}I_i
\end{equation}
and similarly for $L$ and $V$, where $N$ is the number of phase bins in the on-pulse region. The calculation of $L/I$ and $V/I$ was then straightforward, with the error on each measurement determined by
\begin{equation}\label{eqn: linear err}
    \sigma_{\left(L/I\right)} = \frac{L}{I}\times\left(\sqrt{\left(\frac{\sigma_{I}}{I}\right)^2+\left(\frac{\sigma_{L}}{L}\right)^2}\right)\sqrt{N}
\end{equation}
and similarly for $\sigma_{\left(V/I\right)}$.

To calculate $\left|V\right|$, we followed a method analogous to that of \cite{kj04}. We note that their methodology was designed for evaluating $\left|V\right|$ in the case of single pulses. We adapt this method for our fully-integrated profiles, with the caveat that when analysed on a single-pulse basis, each pulsar may show significantly higher $\left|V\right|$ than the pulse-averaged values described here. For each bin in the pulse profile, we produced a bias-corrected estimate of $\left|V\right|_i$, according to
\begin{equation}\label{eqn: abs v bias correction}
    \left|V\right|_i = \left|V_i\right|-\sqrt{\frac{2}{\pi}}\sigma_V.
\end{equation}
We then determined $\sigma_{\left|V\right|}$ as the RMS of the same off-pulse region used to calculate $\sigma_I$, $\sigma_L$ and $\sigma_V$. The final measurement of $\left|V\right|/I$ and its associated uncertainty were then produced in the same manner as $L/I$ and $V/I$, with reference to Equations~\ref{eqn: sumI} and \ref{eqn: linear err}.

\subsection{RVM fitting}\label{subsec: rvm fitting}

A small subset of our \PSRnumPKSFAST pulsars have sufficiently well-defined polarisation angle (PA) curves so as to enable their viewing geometry to be modelled using the Rotating Vector Model \citep[RVM;][]{rc69}. Within this model, the PA of the linearly-polarised emission of each pulsar is a function of the angle $\alpha$ between the rotational and magnetic axes, the angle $\beta$ of closest approach between the line-of-sight and the magnetic axis, and the rotational phase of the pulsar $\phi$. In our analysis, we follow the methodology of \cite{rwj15}, using the \textsc{psrsalsa} software package to determine constraints on the $\alpha$ and $\beta$ parameters for each pulsar. We note that \textsc{psrsalsa} uses the same conventions for $\alpha$ and $\beta$ as used by \cite{rc69} and \cite{rwj15}, while other publications use a modified convention \citep[PSR/IEEE; see e.g.][]{ew01,vmjr10,jk19} wherein $\beta$ has the opposite sign. Care should be taken when comparing and converting between these conventions.

While a simple RVM fit to a pulsar's measured PA swing can provide some constraints on the possible values of $\alpha$ and $\beta$, a more detailed analysis as laid out in \cite{rwj15} can provide significant additional constraint on the viewing geometry. These constraints depend on the width of the open-field-line region $W_\text{open}$, the pulse phase of the inflection point $\phi_{0}$ (at which the line-of-sight passes through the steepest gradient in the PA curve), and the pulse phase of the fiducial point $\phi_\text{fid}$ (the phase at which the line-of-sight passes through the plane containing both the rotational and magnetic axes). These measurements will also depend upon the assumptions made in modeling the emission region of the pulsar and how this model relates to the observed emission profile.

In our analysis, we considered only a simple model of the emission region, where the observed profile is assumed to fully span the open-field-line region. We acknowledge that more complex models of the emission region exist \citep[e.g. the core-cone model, see][]{rankin90,rankin93}, and intend to present a more sophisticated analysis (taking advantage of the full bandwidth of the UWL receiver) utilising these models as part of a future publication. We therefore determined $W_\text{open}$ using the $W_{10}$ values listed in Table~\ref{tab: polarisation table}, with the uncertainty determined using the profile widths at 5\% and 15\% of the pulse peak ($W_{5}$ and $W_{15}$ respectively, again measured with reference to the analytic profile). The inflection point $\phi_{0}$ was determined by \textsc{psrsalsa} as part of the RVM-fitting procedure, which also provided an uncertainty for $\phi_{0}$. The fiducial point $\phi_\text{fid}$ was approximated as the mid-point of the profile according to the positions of the measured $W_{10}$ values. The uncertainty on $\phi_\text{fid}$ was determined using the maximum left- and right-ward offsets allowable for $\phi_\text{fid}$ from the measured positions of $W_{5}$ and $W_{15}$. However, in order to account for the expected relativistic aberration and retardation effects \citep[see e.g.][]{bcw91}, only combinations where $\Delta\phi = \phi_{0} - \phi_\text{fid} \geq 0$ were considered when determining acceptable geometries. We have explicitly indicated in the text where this restriction has been applied.

\section{Summaries of individual pulsars}\label{sec: pulsars of interest}

\subsection{PSR~J0021$-$0909}\label{subsec: C35}

The profile of PSR~J0021$-$0909 consists of a single asymmetric pulse, with a wide left `shoulder' on the pulse indicating a profile which may have multiple underlying components. This hypothesis is bolstered by the observed changes in both $L$ and $V$ across the profile. The modest amount of $L$ in the profile is concentrated on this left `shoulder', and drops away at the same pulse phase as the sign of $V$ is seen to change, with the highest concentration of $V$ aligning with the primary pulse component to the right.

The remaining notable characteristic of PSR~J0021$-$0909 is that it has often proved very difficult to detect during follow-up observations. The pulsar was classed as `undetectable' in the PDFB4 and BPSR observing bands listed in Table~\ref{tab: observing setup} in approximately 58\,\% of observations, with individual integration times as long as 6000\,s. Furthermore, in those observations in which the pulsar \textit{was} detectable, its brightness was often seen to gradually vary in both time and frequency. This, coupled with the pulsar's low DM and the lack of any sudden changes in the pulsar's brightness (which might be indicative of nulling or moding) indicates that the primary cause of these non-detections is likely to be interstellar scintillation. 

\subsection{PSR~J0344$-$0901}\label{subsec: C37}

With a single-component pulse shape and only the weak suggestion of an asymmetric scattering tail, the pulse profile of PSR~J0344$-$0901 comes close to the theoretical Gaussian `ideal'. While the pulsar possesses only modest linear polarisation, it is sufficiently well-defined to allow for a distinct PA swing to be measurable across the profile. However, attempts to model this curve using the RVM indicated that the inflection point $\phi_{0}$ could not be well measured, and was likely to fall outside the region of phase sampled by the pulse profile. Subsequently, we were unable to place meaningful constraints on either the values of $\alpha$ or $\beta$ for this pulsar, and we did not attempt to model its geometry further.

\begin{figure*}
\begin{center}
    \includegraphics[width=\textwidth]{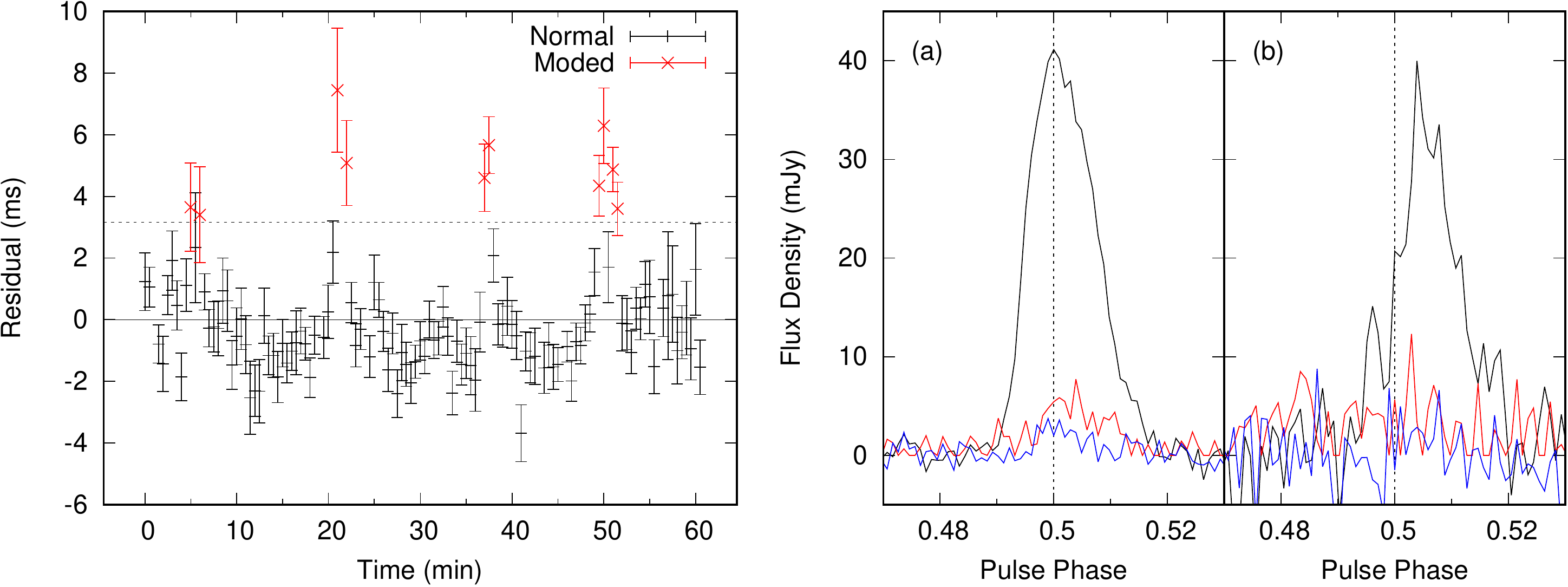}
\end{center}
\caption{Mode-changing  behaviour in a 1-hr observation of PSR~J0344$-$0901 recorded on 2018 October 16th. The left plot shows timing residuals from the observation, as calculated against the timing model described in Table~\ref{tab: timing table 1}. Each TOA was calculated using 30\,s of data, or approximately 25 summed pulses. The dashed horizontal line corresponds to 2.5 times the weighted RMS listed in Table~\ref{tab: timing table 1}. Sub-integrations corresponding to the black TOAs below the line were summed to create the `normal' profile seen in sub-plot (a). Sub-integrations corresponding to the red TOAs above the line were summed to create the `moded' profile seen in sub-plot (b). Both profiles were rotated equally such that the peak of the profile in (a) aligned to a pulse phase of 0.5, indicated by the vertical dashed line.}\label{fig: C37 singlepulse} 
\end{figure*}

PSR~J0344$-$0901 is otherwise notable for its unusual single-pulse behaviour. Typically, single-pulse information on PSR~J0344$-$0901 is unavailable from Parkes data due to the pulsar's low flux density. However, a 1-hr observation on 2018 October 16th caught the pulsar during an unusually-strong up-scintillation, at roughly four times its mean flux density. This allowed for a study of the pulsar's behaviour over much shorter time-scales than had previously been possible. Figure~\ref{fig: C37 singlepulse} shows the TOAs derived from this observation, with one TOA for each of the 30-s sub-integrations used to initially fold the data (each sub-integration contains approximately 25 summed pulses).

It is evident from Figure~\ref{fig: C37 singlepulse} that PSR~J0344$-$0901 shows a type of mode-changing behaviour. The transitions between the `normal' and `moded' states are not sharp from one pulse to the next, but appear to take place gradually over the course of many tens of seconds. Similarly, the duration of each `moded' state has only been seen to last between 1 to 2 minutes. In order to analyse the different states of PSR~J0344$-$0901 discretely, we have chosen a residual threshold of 2.5 times the weighted RMS listed in Table~\ref{tab: timing table 1} (roughly 3.2\,ms) to distinguish between the two states, which is shown in sub-plot (a) of Figure~\ref{fig: C37 singlepulse}. As can be seen in sub-plot (b) of Figure~\ref{fig: C37 singlepulse}, the moded profile is characterised by a lag in the pulse-peak of approximately 6\,ms (or approximately 0.005 units of pulse phase), as well a potential reduction in flux density. However, given the limited number of integrated pulses in sub-plot (b) (only $\sim250$ pulses), the true flux density of the pulsar in its moded state is currently difficult to quantify.

It is therefore possible that PSR~J0344$-$0901 may represent a similar class of object to PSR~B0919$+$06 and PSR~B1859$+$07 \citep{rrw06}. Both of these pulsars are seen to display similar gradual transitions between their normal and moded states (often referred to as `swooshes'), with the notable difference being that while PSR~J0344$-$0901's moded pulses are seen to lag, the moded pulses of PSR~B0919$+$06 and PSR~B1859$+$07 arrive early. In both of these previously-known pulsars, a correlation has been found between their changing emission states and their spin-down rates \citep{psw+15,psw+16}, however it is too early to determine whether such a correlation exists within PSR~J0344$-$0901.

We remark finally upon the apparent periodicity seen in Figure~\ref{fig: C37 singlepulse}, wherein the moding events in PSR~J0344$-$0901 appear to be regularly spaced by approximately 16\,min intervals. Similar periodicities have also been identified in PSR~B0919$+$06 and PSR~B1859$+$07 \citep{worw16}, with the suggestion that preivously-undetected orbital motion may be responsible. This hypothesis was initially suspected in the case of PSR~J0344$-$0901, however subsequent observations with FAST appear to have ruled out this scenario. In two 1-hr observations, moding events were observed by FAST but with no apparent periodicity, while in a third 1-hr observation, no moding events were observed at all. These observations are part of an ongoing study of PSR~J0344$-$0901 with the FAST telescope, and be reported on in greater detail in a future publication.

\subsection{PSR~J0529$-$0715}\label{subsec: C29}

PSR~J0529$-$0715 displays one of the more complex profiles of the pulsars presented in this paper. In the development of an analytical standard for this pulsar, we employed four individual Gaussian components, one for the left and right `shoulders' and one for each of the central pulse peaks. However, despite averaging over approximately $1.68\times10^{5}$ pulses, it remains uncertain whether each of these features are genuine, or whether the profile is still contaminated by noise. An analysis of the profile over time indicates that the shoulders remain relatively consistent between observations, but that the central peaks may in fact be a single peak of emission. Furthermore, with only a moderate amount of $L$ and only scattered measures of PA across the profile, it is difficult to determine how well this analytical model maps to the true structure of the pulsar's emission region. Higher S/N observations will assist in clarifying these ambiguities.

\subsection{PSR~J0803$-$0942}\label{subsec: C36}

As seen in Figure~\ref{fig: pulse profiles 1} and Table~\ref{tab: polarisation table}, the standout feature of PSR~J0803$-$0942 is its extremely high fraction of linearly polarised flux, with its $L/I$ measured at approximately 96\%. Although this is by far the highest $L/I$ seen amongst the \PSRnumPKSFAST pulsars presented here, it is not an unknown feature amongst the general population of pulsars whose flux densities and polarisation characteristics have been measured at 1.4\,GHz \citep[see e.g.][]{jk18}. This high $L/I$ also allows for the measurement of PA across the entire pulse profile. However, the trend in PA appears to be almost flat, rising only slightly across the profile. Consequently, as with PSR~J0344$-$0901 we are unable to provide useful constraints on $\phi_{0}$, $\alpha$ or $\beta$ for this pulsar, and we did not attempt to model its geometry further. Otherwise, PSR~J0803$-$0942 has only minimal circular polarisation, and displays a nearly-symmetrical single-component profile, with the exception of a weak emission feature on the profile's rising edge.

\subsection{PSR~J1851$-$0633}\label{subsec: C16}

The profile of PSR~J1851$-$0633 contains two distinct components. With the current sensitivity available from Parkes, there is no detectable intermediate `bridging' emission between these two components. The pulsar is moderately linearly polarised across both profile components, although it is presently only possible to obtain measurements of PA in the left component. A higher S/N profile may also allow for PA measurements in the right component, after which an RVM fit may be possible. 

\begin{figure}
\begin{center}
\includegraphics[width=\columnwidth]{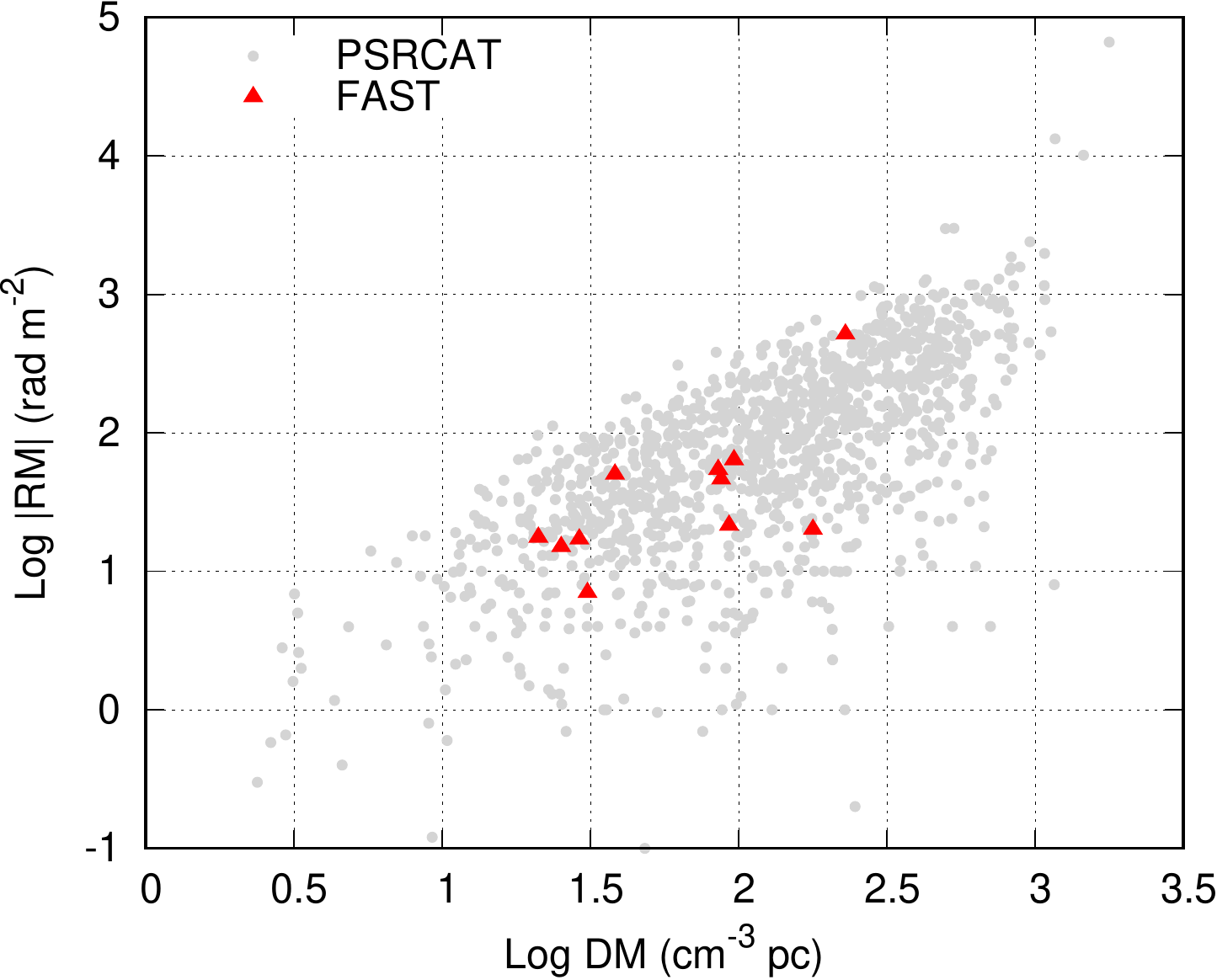}
\end{center}
\caption{Positions of the \PSRnumPKSFAST newly-discovered pulsars (red triangles) within the DM-RM distribution of the known pulsar population (grey circles), consisting of \PSRnumBACKGROUNDrmdm pulsars for which \textsc{psrcat} \citep[version 1.61;][]{psrcat} lists well-measured values of both DM and RM.}\label{fig: rm-dm} 
\end{figure}

PSR~J1851$-$0633 is also notable in that, as listed in Table~\ref{tab: polarisation table}, it possesses the highest RM of any of the pulsars listed in this paper by a significant margin. However, given the general correlation between increasing $\left|\text{RM}\right|$ and increasing DM as seen in Figure~\ref{fig: rm-dm}, and the fact that PSR~J1851$-$0633's DM is also the largest of the pulsars presented in this paper, this large RM is not especially surprising. As shown in Figure~\ref{fig: rm-dm}, its DM and RM values are well within the existing distribution, as are those of the other pulsars featured in this paper.

It is also worth briefly remarking on the fact that FAST was able to detect this higher-DM pulsar despite its search at lower radio frequencies. The pulse-smearing effect of interstellar scattering, which degrades pulsar detectability, is known to increase both with increasing DM and decreasing frequency \cite[see e.g.][]{brc+04}. However, PSR~J1851$-$0633 was initially detected in a sub-banded search between 500 and 750\,MHz, avoiding the worst of the scattering at the lowest end of the UWB frequency band.

\subsection{PSR~J1900$-$0134}\label{subsec: C4}

As with PSR~J1851$-$0633, PSR~J1900$-$0134 also shows a clearly double-peaked profile. However, unlike PSR~J1851$-$0633 there does appear to be a weak `bridge' of emission between the two primary components. Each of these components shows a moderate fraction of linearly polarised flux, along with a weaker yet significant fraction of circularly polarised flux. Also of note is PSR~J1900$-$0134's characteristic age; it is the youngest pulsar presented in this paper, with a $\tau_\text{c}$ of only 0.95\,Myr.

\begin{figure}
\begin{center}
    \includegraphics[height=\columnwidth, angle=270]{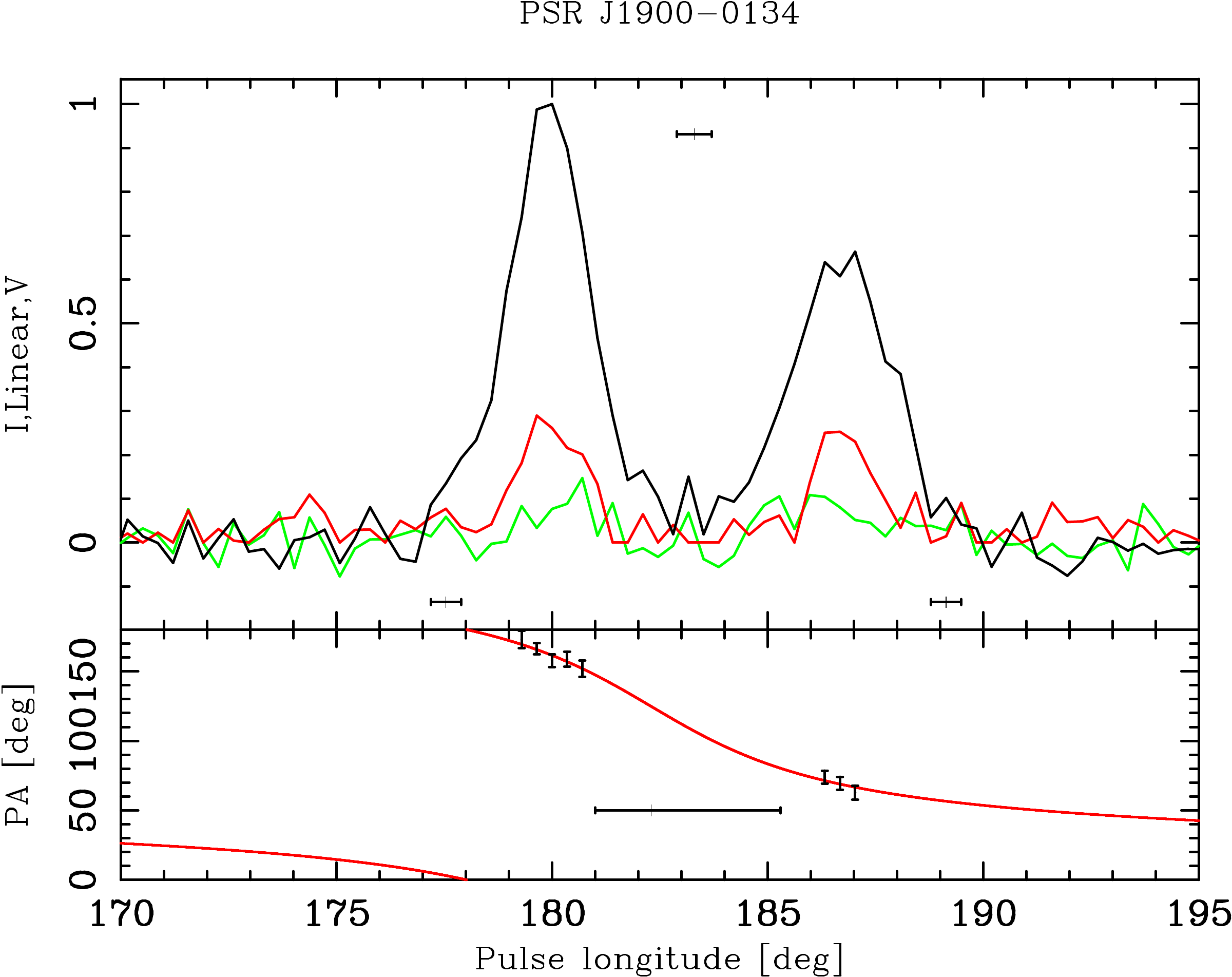}
    \vspace{0.5cm}
    \includegraphics[height=\columnwidth, angle=270]{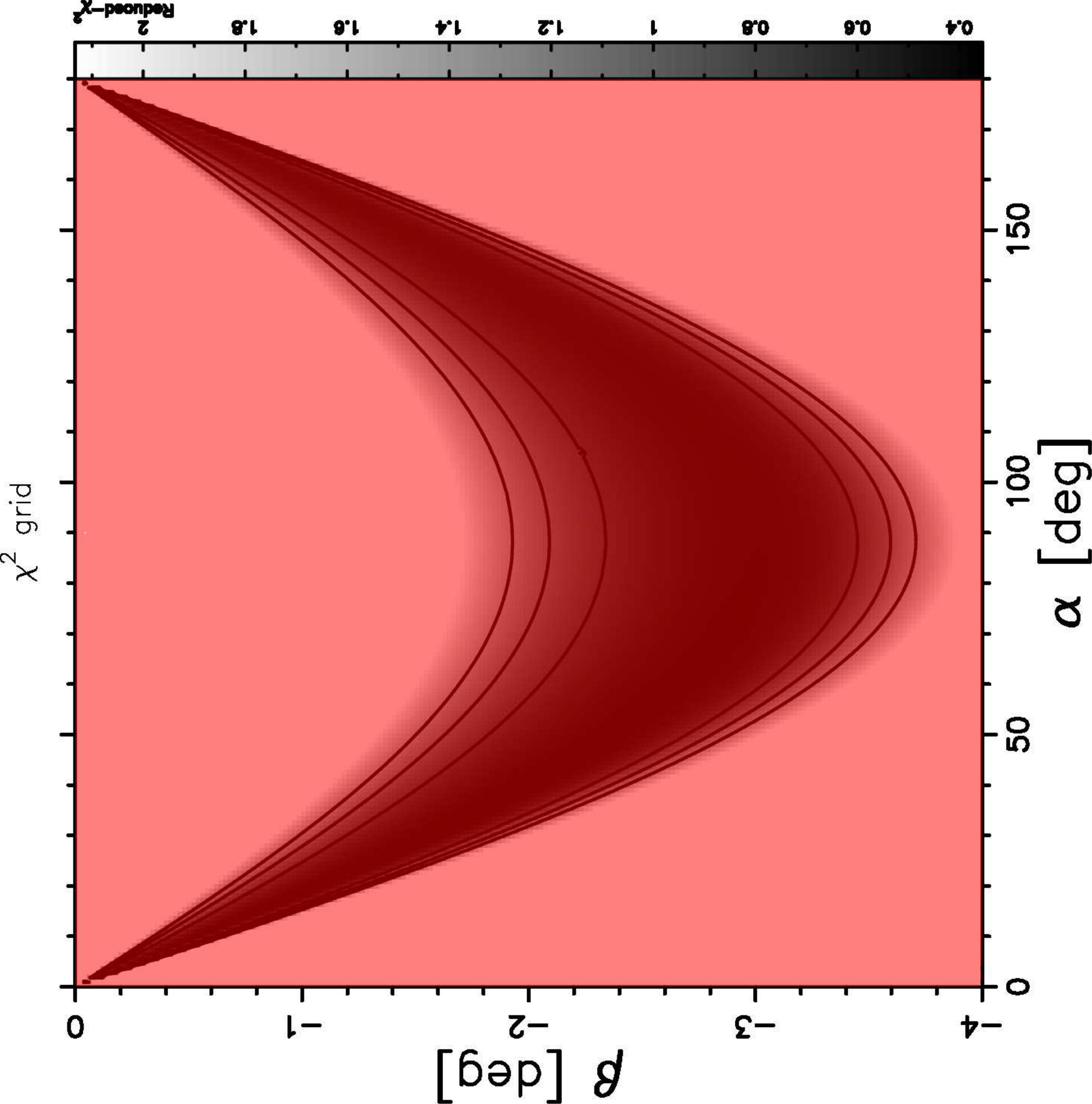}
\end{center}
\caption{Constraints on the RVM-fit and modelled geometry of PSR~J1900$-$0134. The upper plot shows the pulsar's polarisation profile, with normalised total intensity $I$ in black, linear polarisation $L$ in red and circular polarisation $V$ in green. The horizontal error bar above the profile shows the position of the fiducial point $\phi_\text{fid}$, while those below the profile show the points used to determine the profile width $W_\text{open}$. Measurements of PA are shown below the profile, along with a best-fit RVM solution in red, corresponding to $\alpha=7.0^\circ$ and $\beta=-0.4^\circ$ (as per our chosen convention for $\alpha$ and $\beta$, see Section~\ref{subsec: rvm fitting}). The horizontal error bar shows the position of $\phi_{0}$, which maps to the RVM solution at $\text{PA} = -55.2^\circ$. The lower plot shows a $\chi^{2}$ map of $\alpha$ and $\beta$ following the RVM-fit to the PA values, with outwards contours corresponding to 1, 2 and 3-$\sigma$ respectively. The region in red (here covering the entire lower plot, indicating a lack of constraint) shows the allowed geometry based upon $W_\text{open}$, $\phi_{0}$ and $\phi_\text{fid}$, as per the methodology laid out in Section~\ref{subsec: rvm fitting}.}\label{fig: C4 rvm}
\end{figure}

As seen in Figure~\ref{fig: pulse profiles 1}, PSR~J1900$-$0134 also has well-constrained PA measurements across each of its primary components. We have therefore attempted to model the geometry of this pulsar by applying the RVM model as per the methodology laid out in Section~\ref{subsec: rvm fitting}. The resulting fit is shown in Figure~\ref{fig: C4 rvm}. Assuming that the visible profile fully spans the open-field-line region, we measured a pulse width of $W_\text{open} = {12.0^\circ}{^{+0.7}_{-0.7}}$ with a fiducial phase chosen at the midpoint of the profile of $\phi_\text{fid}={183.3^\circ}{^{+0.4}_{-0.4}}$. Unfortunately, it seems likely that the inflection phase of the PA swing $\phi_{0}$ lies within the region of weak bridging emission where we are unable to directly measure values of PA, leaving its position uncertain. The best estimate provided by \textsc{psrsalsa} is that $\phi_{0}={182.3^\circ}{^{+3.0}_{-1.3}}$. Remembering the imposed restriction that $\Delta\phi\geq0$, this results in $\Delta\phi={0.0^\circ}{^{+2.3}_{-0.0}}$.

From the results of the fit shown in Figure~\ref{fig: C4 rvm}, it seems clear that the geometry of PSR~J1900$-$0134 can only be poorly constrained. While the impact angle is restricted to within $-4^\circ<\beta<0^\circ$, the magnetic inclination angle $\alpha$ is essentially unconstrained. Attempting to further limit the geometry of the pulsar using our measured $W_\text{open}$ and $\Delta\phi$ results in no additional constraint due to the large uncertainty in $\phi_{0}$. With reference to Equation 3 in \cite{rwj15}, our measured $\Delta\phi$ would imply an upper limit on the emission height of $h_\text{em} \lesssim 880\,\text{km}$. We note that taking the nominal values of $\phi_{0}$ and $\phi_\text{fid}$ would result in $\Delta\phi<0$, which may suggest that the detected profile does not fully fill the open-field-line region. However, this result may also be due to the large uncertainty in $\phi_{0}$.

\subsection{PSR~J1919$+$2621}\label{subsec: C7}

PSR~J1919$+$2621 presents a single-component, asymmetric, slightly-triangular pulse profile. The profile contains a moderate amount of linear polarisation, with little to no circular polarisation. Perhaps the most striking feature comes from the unusual trend seen in the pulsar's PA distribution. The PA trend experiences a sharp transition approximately coinciding with the profile peak, at which the sign of the slope changes value and a discrete jump in PA is observed. Due to this change in slope, and the fact that the jump is only very small (i.e. not close to $90^\circ$), it seems unlikely that this transition is due to an orthogonal polarisation mode. Nor does the transition appear to directly coincide with any immediate features in the linear or circular polarisation profiles. We therefore tentatively suggest that this transition indicates the presence of two over-lapping emission regions within the open-field-line region of the pulsar, each with its own distinct polarisation properties. This is not the first time such a non-orthogonal PA jump has been seen \citep[see e.g.][]{dhm+15,jk18}, and we will continue to investigate its nature with ongoing observations.

\subsection{PSR~J1926$-$0652}\label{subsec: C12}

With the widest fractional duty cycle of any of the pulsar presented here, PSR~J1926$-$0652 displays two dominant, well-separated pulse components, linked by an intermediate bridge of pulsed emission. The pulsar's integrated profile displays substantial linear polarisation, comprising nearly half of the total flux density, allowing for the measurement of a well-defined, characteristic PA swing across the profile. Further details regarding RVM-fitting and the geometry of PSR~J1926$-$0652, as well its rich single-pulse behaviour, can be found in \cite{zlh+19}.

\subsection{PSR~J1931$-$0144}\label{subsec: C6}

The profile of PSR~J1931$-$0144 is perhaps the most unremarkable of the pulsars presented in this paper, with a broad, single-component Gaussian pulse. Although the profile shows moderate linear polarisation, the low S/N of the profile prevents the measurement of more than a handful of PA values. The few PA data points shown in Figure~\ref{fig: pulse profiles 2} hint at a more pronounced PA swing, which may be detectable with additional, more-sensitive observations. PSR~J1931$-$0144 is otherwise notable for its low $\dot{P}$ as listed in Table~\ref{tab: timing table 3}, placing it towards the bottom of the `normal' pulsar distribution in the $P$-$\dot{P}$ diagram in Figure~\ref{fig: discovery p-pdot} and giving it the largest characteristic age of the reported pulsars, with $\tau_\text{c}=320\,\text{Myr}$.

\subsection{PSR~J1945$+$1211}\label{subsec: C14}

The profile of PSR~J1945$+$1211 appears almost as a mirror image of PSR~J1926$-$0652, as it also possesses two clear peaks linked by a bridge of intermediate emission, but with the left peak being stronger than the right, the inverse of PSR~J1926$-$0652. Also similar to PSR~J1926$-$0652, roughly half of the total flux density of the pulsar is linearly polarised, again allowing for the measurement of a well-defined PA swing across almost the entire pulse profile. We have therefore attempted to model the geometry of this pulsar by applying the RVM model as per the methodology laid out in Section~\ref{subsec: rvm fitting}. 

\begin{figure}
\begin{center}
    \includegraphics[height=\columnwidth, angle=270]{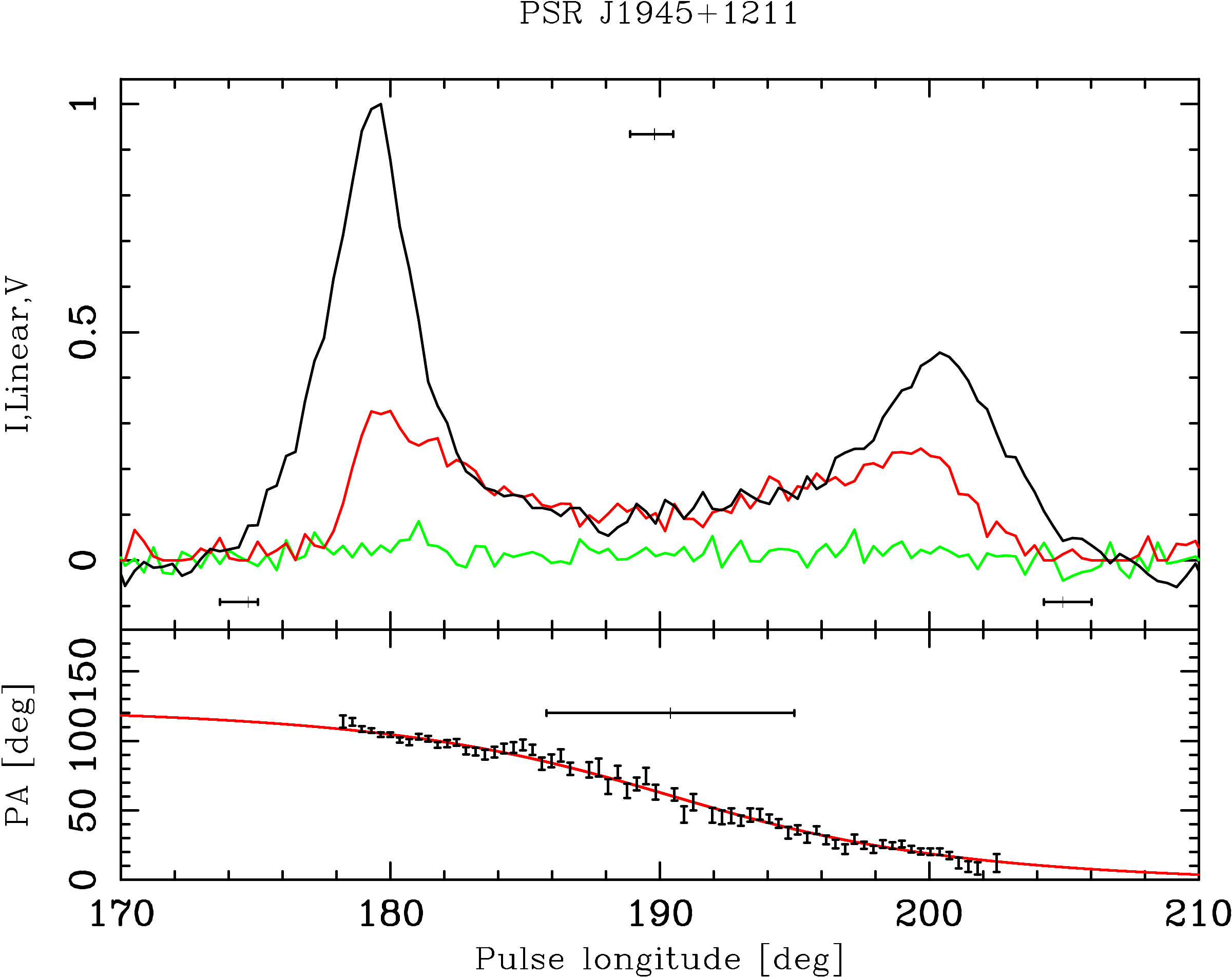}
    \vspace{0.5cm}
    \includegraphics[height=\columnwidth, angle=270]{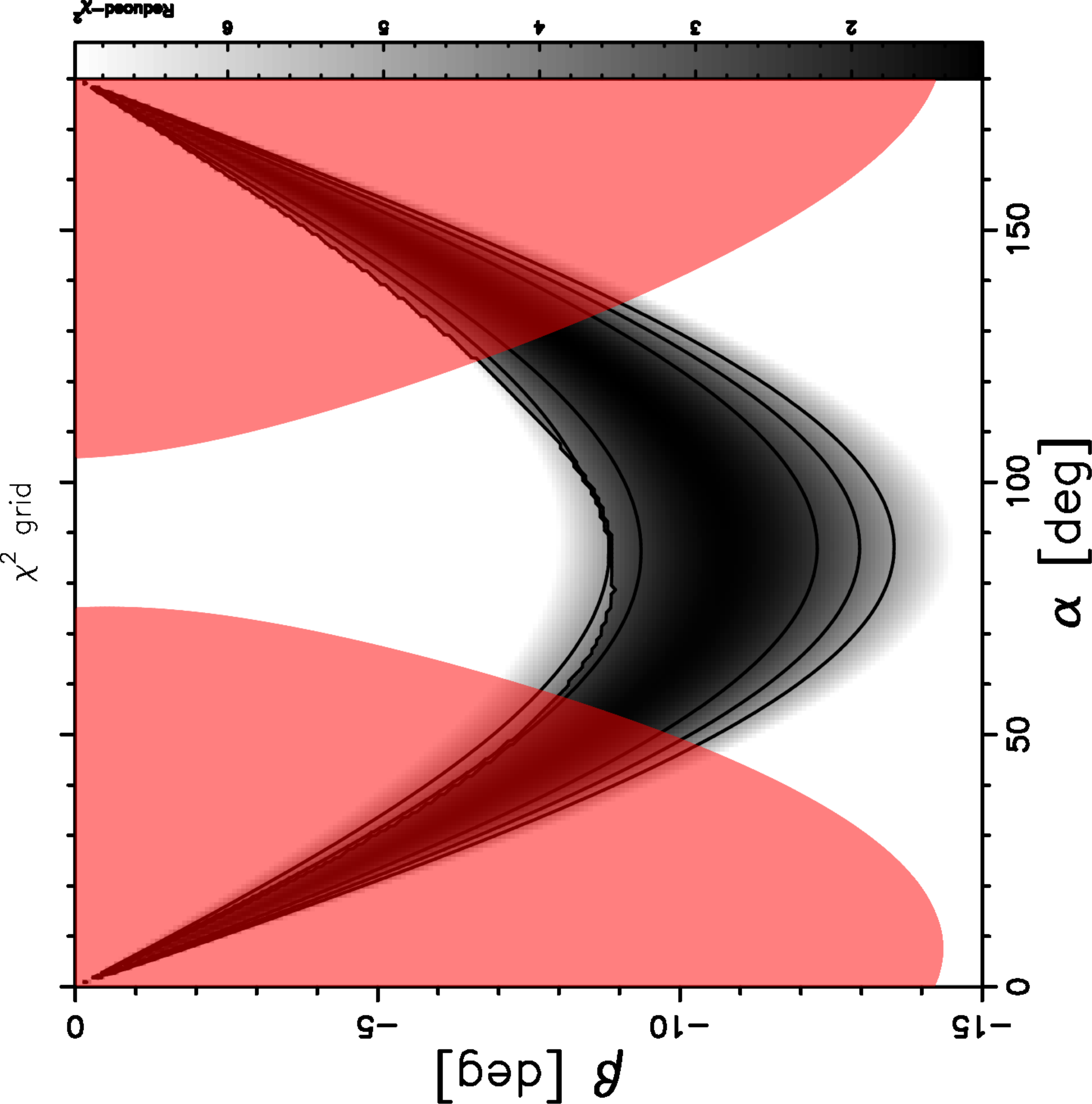}
\end{center}
\caption{Constraints on the RVM-fit and modelled geometry of PSR~J1945$+$1211. Details as per Figure~\ref{fig: C4 rvm}. The best-fit RVM solution corresponds to $\alpha=142.5^\circ$ and $\beta=-6.1^\circ$. $\phi_{0}$ maps to the RVM solution at $\text{PA} = 60.8^\circ$.}\label{fig: C14 rvm}
\end{figure}

The resulting RVM-fit is shown in Figure~\ref{fig: C14 rvm}. Assuming that the visible profile fully spans the open-field-line region, we measured a pulse width of $W_\text{open} = {30.6^\circ}{^{+2.1}_{-1.1}}$ with a fiducial phase chosen at the midpoint of the profile of $\phi_\text{fid}={189.8^\circ}{^{+0.7}_{-0.9}}$. Unlike the case of PSR~J1900$-$0134, our measured PA swing appears to directly sample the inflection point $\phi_{0}$, however the gentle slope of the swing still leaves significant uncertainty in its position. The best estimate provided by \textsc{psrsalsa} is that $\phi_{0}={190.4^\circ}{^{+4.6}_{-4.6}}$ to within 1-$\sigma$ of uncertainty. Remembering the imposed restriction that $\Delta\phi\geq0$, this results in $\Delta\phi={0.5^\circ}{^{+5.7}_{-0.0}}$.

Based upon the results shown in Figure~\ref{fig: C14 rvm}, we are able to place some tentative constraints on the geometry of PSR~J1945$+$1211. Based only upon the RVM-fit to the PA swing, the impact parameter $\beta$ can be constrained within approximately $-14^\circ<\beta<0^\circ$, while the magnetic inclination angle $\alpha$ is essentially unconstrained. However, combining this with limits imposed by our measured $W_\text{open}$ and $\Delta\phi$ further restricts the parameters to within approximately $-10^\circ<\beta<0^\circ$ and either $\alpha < 60^\circ$ or $\alpha>125^\circ$, implying that the rotational and magnetic axes of PSR~J1945$+$1211 are more likely to be aligned than orthogonal. Again with reference to Equation 3 in \cite{rwj15}, our measured $\Delta\phi$ would imply an upper limit on the emission height of $h_\text{em} \lesssim 6100\,\text{km}$. The uncertainty in the position of $\phi_{0}$ is perhaps the largest limiting factor in further constraining the geometry of this pulsar. Therefore, future observations should focus on improving the S/N of the measured profile, especially through the region of bridging emission, as well as to characterise the profile over a wider bandwidth in order to truly understand the extent of the emission region.

One final note of interest regarding PSR~J1945$+$1211 is that it was frequently seen to display significant variability over very short timescales, typically from one sub-integration to the next. While some of this may be attributable to scintillation, it seems likely that given the pulsar's long period and the typical 30-s duration of each sub-integration, this is simply the result of the pulsar's intrinsic pulse-to-pulse variability, with each sub-integration only including approximately 6 to 7 pulses. Unfortunately, the single pulses of PSR~J1945$+$1211 largely fall below the threshold of Parkes' sensitivity, but future dedicated followup by FAST may be able to probe this aspect of PSR~J1945$+$1211's behaviour.

\subsection{PSR~J2323$+$1214}\label{subsec: C13}

PSR~J2323$+$1214 displays two clear pulse components, however it is unclear as to whether there is any bridging emission linking the two. Although the profile shows moderate $L/I$, we are unable to detect any clear trend in the PA across the profile. As with PSR~J1945$+$1211, PSR~J2323$+$1214 also shows similar short-timescale variability. We believe that this is once more a combination of the effects of scintillation with the pulsar's long period, which only allows for approximately 8 pulses per 30-s sub-integration. The single pulses of PSR~J2323$+$1214 also largely fall below the detection threshold of Parkes, and so we once again recommend additional followup by FAST.

\section{Discussion}\label{sec: discussion}

\subsection{Detectability in earlier surveys}\label{subsec: earlier surveys}

While these \PSRnumPKSFAST pulsars were discovered through the use of FAST and the significant increase in gain afforded by such a large radio telescope, it is worth considering whether any of them could have been discovered by earlier pulsar surveys. Since we have characterised the behaviour of these pulsars in the 20-cm observing band, we consider here only a set of major pulsar surveys undertaken at this frequency which have been conducted over approximately the previous two decades, and note that this overview is unlikely to be fully comprehensive. We do not intend this as a review of the capabilities of FAST, since the initial discovery of these pulsars by FAST was at much lower frequencies where their properties have yet to be fully evaluated.

We first consider those surveys undertaken with the Parkes 64-m radio telescope. Both PSRs~J1851$-$0633 and J1900$-$0134 fall within the sky region of the Parkes Multibeam Pulsar Survey \citep[PMPS;][]{mlc+01}, for which the average limiting flux density is approximately 0.22\,mJy. We then adjust this value to account for both the narrow pulse widths of these pulsars and the specific sky temperature\footnote{Derived from \cite{rdb+15}.} at each position (which \citealt{mlc+01} did not take into account). However, even after these corrections, the flux densities of both pulsars appear to fall below the detectable threshold of the PMPS. Therefore, it is unsurprising that neither of these pulsars were previously discovered by the PMPS.

PSR~J0803$-$0942 lies within the boundaries of the Parkes High-Latitude pulsar survey \citep[PKSHL;][]{bjd+06}. For a pulsar with a duty cycle of 1\,\% (comparable with 1.4\,\% for PSR~J0803$-$0942) and located at the center of the central beam of the MB20 receiver, \cite{bjd+06} reports a limiting flux density of $\sim0.2\,\text{mJy}$. This is already a factor of two larger than the measured flux density of PSR~J0803$-$0942, and so the non-detection of this pulsar in PKSHL is also to have been expected.

Considering the modern generation of Parkes MB20 pulsar surveys, we next turn to the HTRU-S survey \citep{kjvs10}. The only pulsar to fall within either the HTRU-S LowLat or HTRU-S MedLat sky regions was PSR~J1851$-$0633, which was previously detected by the HTRU-S LowLat survey and represents a shared discovery. Meanwhile, all but three of our reported pulsars (namely PSRs~J1919$+$2621, J1945$+$1211 and J2323$+$1214) fall within the HTRU-S High Latitude survey \citep[HTRU-S HiLat;][]{kjvs10}, which has a limiting flux density of approximately 0.27\,mJy. The only pulsar within the HTRU-S HiLat region with a flux density above this value is PSR~J1926$-$0652. However, given the reported nulling behaviour of this pulsar \citep{zlh+19} and the fact that each HiLat observation is only 4.5 minutes long, it is unsurprising that this pulsar eluded detection in this earlier survey, as it may easily have been observed during its `off' state.

One additional modern pulsar survey undertaken at Parkes has been the Survey for Pulsars and Extragalactic Radio Bursts \citep[SUPERB;][]{kbj+18}. In many ways, SUPERB represents a more-sensitive successor to HTRU-S HiLat, surveying much of the same region of sky but with just over double the integration length, bringing down its limiting flux density to approximately 0.15\,mJy. Five of our reported pulsars (PSRs~J0803$-$0942, J1851$-$0633, J1900$-$0134, J1926$-$0652 and J1931$-$0144) fall within the SUPERB region, with three of these (PSRs~J1900$-$0134, J1926$-$0652 and J1931$-$0144) having flux densities above the SUPERB flux density limit. For each of these three pulsars, we retrieved all SUPERB data publicly available via the ATNF Pulsar Data Access Portal\footnote{https://data.csiro.au/dap/public/atnf/pulsarSearch.zul} which was within 7 arcminutes of their timed positions, and folded each observation according to the ephemerides in Tables~\ref{tab: timing table 1}, \ref{tab: timing table 2} and \ref{tab: timing table 3}. Only PSR~J1931$-$0144 was detectable, but at a typical folded S/N between 7 and 7.5, well below the SUPERB detection threshold. 

Lastly, we consider major L-band surveys undertaken by radio telescopes other than Parkes. One such survey is the Arecibo L-band Feed Array pulsar survey \citep[PALFA;][]{cfl+06}, which was undertaken with the 300-m Arecibo telescope in Puerto Rico. This survey observed a region of the Galactic plane between declinations of $-1^\circ<\delta<+38^\circ$ and Galactic latitudes of $\left|l\right|<5^\circ$, which unfortunately\footnote{Or fortunately, depending on one's perspective.} excludes all pulsars reported in this paper, and so we make no further comment. Finally, Effelsberg has also been conducting the HTRU-North survey \citep[HTRU-N;][]{bck+13}, a northern counterpart to the HTRU-S survey. As an all-sky survey covering declinations of $\delta>-20^\circ$, the HTRU-N region includes all \PSRnumPKSFAST pulsars reported here, and is theoretically sensitive enough to have detected approximately half of them with a standard periodicity search. However, as large portions of the survey remain to be observed and/or searched, it may simply be down to chance that HTRU-N was not able to discover these pulsars prior to their detection by FAST.

\subsection{Non-detections of remaining candidates}\label{subsec: non-detections}

We note that Parkes was unable to confirm \PSRnumFASTnonconfPKS pulsar candidates from FAST despite having observed them, often multiple times. However, at this stage we can not draw any conclusions regarding the validity of the candidates based upon these non-detections. This is due to the large number of potential contributing factors, which are neither easily separable nor easily quantifiable. Among others, these factors include:
\begin{itemize}
\item \textbf{Scintillation:} The effects of interstellar scintillation are typically more pronounced for narrow-bandwidth observations at lower values of DM \citep[see e.g.][]{sutton71, backer75}. Of the \PSRnumFASTcandsPKS FAST candidates in the Parkes sky, 24 have DMs less than $100\,\text{cm}^{-3}\,\text{pc}$, and 13 have DMs less than $50\,\text{cm}^{-3}\,\text{pc}$. Therefore, it is likely that scintillation may have played a role in the non-detection of a number of these pulsars.
\item \textbf{Spectral index:} The flux density of pulsars typically decreases as observing frequency increases, with most pulsars displaying steep spectral indices \citep[see e.g.,][]{jvk+18}. As these pulsar candidates were discovered at observing frequencies less than $\sim800\,\text{MHz}$, this significantly increases the difficulty of confirming them with both the $\sim1.4\,\text{GHz}$ MB20 receiver and even the UWL, whose observing band starts at $\sim700\,\text{MHz}$.
\item \textbf{Sensitivity:} Even during its commissioning phase, the gain of FAST during its early pulsar observations was estimated to be $10.1\,\text{K\,Jy}^{-1}$ \citep{qpl+19}, nearly fourteen times larger than Parkes' gain of $0.74\,\text{K\,Jy}^{-1}$. In order to compensate, integration times on Parkes typically need to be larger by often two orders of magnitude. For example, even excluding all other effects listed in this section, a pulsar detected by FAST's wideband receiver in 60\,s with a signal to noise (S/N) of 10 would require at least 2500\,s of integration time using the MB20 with Parkes. These long observation times also have the potential to bias against the detection of compact binary systems \citep[see e.g.,][]{ncb15}.
\item \textbf{Positional errors:} Due to the fact that FAST was still operating in a commissioning and testing phase, errors in position often resulted during early observations. This was exacerbated by the fact that FAST was operating in a driftscan mode (as opposed to a tracking mode), and so the on-sky position of the telescope was constantly changing. Observers at Effelsberg determined in at least one instance that the true position of a confirmed pulsar was separated from its reported candidate position by a significant fraction of a degree, which is likely to have hampered our confirmation efforts (Cruces et~al. in prep.).
\end{itemize}

\subsection{Future wideband analysis}\label{subsec: future wideband}

As noted earlier, we have also recorded a significant quantity of wideband data for each of the \PSRnumPKSFAST pulsars, spanning the full bandwidth of the UWL. We have chosen not to include this data as part of our analysis in this paper for two primary reasons. Firstly, as the UWL is a new instrument and produces significantly larger volumes of data than earlier receivers, we have only recently mastered the data reduction techniques required to efficiently process the full receiver bandwidth. Secondly, as the UWL only came into use part-way through our observing campaign, we wish to collect more data in order to improve the integrated S/N of each pulsar beyond that which was available at the time of publication. 

It is already clear that many of the pulsars reported here will benefit from an analysis utilising the full UWL band. For example, it should be possible to further characterise the properties of the scintillation observed in PSR~J0021$-$0909, as well as to make more reliable detections of the pulsar by analysing portions of the band where the pulsar is scintillating up. A wideband analysis will also be useful in better characterising the profile structure and emission region of many of these pulsars by indicating how the components and polarisation properties change with frequency. Such an analysis will be especially useful in refining our preliminary attempts at modeling the geometry of these pulsars (notably PSRs~J1900$-$0134 and J1945$+$1211) and may allow for the modeling of additional pulsars (e.g. PSRs~J0344$-$0901 and J1851$-$0633).

\subsection{Lessons for future pulsar surveys}

Our experience in using Parkes to follow-up discoveries from FAST has shed light on several topics which are likely to be relevant to the next generation of pulsar surveys to be undertaken by telescopes such as MeerKAT, the Square Kilometre Array (SKA) and future surveys by FAST itself. Particularly in the case of the SKA \citep[see e.g.][]{kbk+15}, much consideration has been given to the problem of efficiently producing well-curated sets of pulsar candidates from increasingly large data volumes. The general consensus has been to move from offline to real-time candidate processing \citep[see e.g.][]{kbj+18} and from manual, human-driven candidate review to automatic systems based upon machine learning techniques \citep[for a recent review, see][]{lsc+16}.

Less discussion is available in the literature on the topic of how these pulsar candidates should be followed up once candidates have been determined, with notable entries including \cite{ckl+04} and \cite{kbk+15}. In the case of just the proposed SKA1-MID, \cite{kbk+15} anticipate the detection of $\sim10,000$ pulsars, more than tripling the current number of known pulsars listed in \textsc{psrcat}\footnote{https://www.atnf.csiro.au/research/pulsar/psrcat/}. At this stage, it is unclear how much time will be available on these next-generation telescopes to follow-up each pulsar candidate once their surveys are complete, with optimisation strategies already under consideration \citep[e.g. in the case of SKA1-MID, using sub-arrays to observe multiple fields simultaneously, see][]{ckl+04}. It is therefore conceivable that many of these surveys will (as per the example of FAST and Parkes) also need to make use of additional, less-sensitive telescopes in order to conduct both candidate confirmation and the long-term timing required to determine phase-connected models of each pulsar's behaviour.

The challenges in such an endeavour would not be insignificant. As highlighted in Section~\ref{subsec: non-detections}, the differences in gain, etc. between current and next-generation telescopes imply that such follow-up observations will require a significantly greater amount of observing time in order to achieve a suitable S/N for each pulsar so as to reliably time the pulsar. This may require additional triaging or prioritising of pulsar candidates beyond the initial candidate identification process, in order to ensure the most efficient use of resources, most crucially the available observing time across all participating telescopes. For example, brighter candidates may be prioritised for follow-up on a current-generation telescope such as Parkes while fainter candidates may be retained for follow-up by the original telescope. Alternatively, candidates could be prioritised based upon scientific merit, with critical discoveries followed up by the most sensitive telescopes while less remarkable pulsars are passed to less sensitive telescopes.

The role of next-generation receivers, including phased-array feeds or wideband receivers, should also be considered when assessing an efficient follow-up strategy. For example, while a telescope such as MeerKAT or FAST may have significantly higher gain at a specific frequency than a telescope such as Parkes, the Parkes UWL receiver may provide an advantage at an alternative frequency where a given pulsar can be more favorably observed (e.g. due to spectral index or interstellar scattering). Similarly, should a small collection of candidates be spatially-coincident on the sky, a telescope equipped with a phased-array feed may be able to observe and fold each of these candidates simultaneously, a technique previously demonstrated by \cite{dch+17}.

The problem of candidate follow-up optimisation is exacerbated in the case of MSPs and binary pulsars, as explored previously by \cite{kbk+15}. These targets typically require a higher cadence in order to determine their orbital and rotational properties. This would have a multiplying effect on the additional resources required to follow-up such pulsars on current generation telescopes. Additionally, as also noted in Section~\ref{subsec: non-detections}, the longer integration times may prevent the confirmation of pulsars in sufficiently tight binaries when considering the application of standard acceleration-based search techniques \citep{ncb15}. This may necessitate that pulsars suspected of being in such a binary system be initially followed up by more sensitive telescopes which can employ shorter integration times, at least until an orbital solution can be determined. Alternatively, confirmations could be made with current generation telescopes through the use of more advanced binary search techniques, such as `jerk' searches \citep[see e.g.][]{ar18,and+19} or orbital deconvolution techniques.

\section{Conclusions}\label{sec: conclusions}
As a pilot program for the upcoming CRAFTS project, we present here \PSRnumPKSFAST pulsars discovered by the FAST UWB in drift-scan mode. They represent all sources from the FAST UWB sample that can be confirmed and well timed by Parkes. Timing models of each pulsar's spin properties and detailed evaluation of their emission properties, including flux calibration and polarization properties, are obtained. Their spin periods range from 0.57 s to 4.76 s. Their characteristic ages range from 0.65 to 320 Myr. They largely reside in the 'normal' pulsar region on the $P$-$\dot{P}$ diagram.

Many display emission features of significant scientific interest, which warrant ongoing follow-up study. At least two pulsars exhibit particular variability in their pulsations. PSR~J0344$-$0901 shows moding `swooshes', which has only been found in a small, but well-documented class of pulsars. Future study of this pulsar will benefit greatly from high-sensitivity, single-pulse studies by FAST, as would PSRs~J1945$+$1211 and J2323$+$1214, which may exhibit their own rich single-pulse behaviour. The other pulsar of note is PSR~J1926$-$0652, which shows a plethora of emission phenomena, including nulling and sub-pulse drifting, which have been presented as part of an earlier paper \citep{zlh+19}.

RVM modeling of the emission geometry was carried out as part of this paper for PSRs~J1900$-$0134 and J1945$+$1211. This is in addition to PSR~J1926$-$0652, which was modeled as part of \cite{zlh+19}. PSR~J1900$-$0134, in particular, can be relatively well constrained to have a pulse width of $W_\text{open} = {12.0^\circ}{^{+0.7}_{-0.7}}$ and an emission height $h_\text{em} \lesssim 880\,\text{km}$. The PA curves of PSRs~J0344$-$0901, J1851$-$0633 and J1931$-$0144 are also promising, although not well modeled by our current analysis. Observations with a much wider bandwidth should better reveal the structure of emission regions as well as their spectral index and scintillation properties.

We close this paper by looking forward to the future of pulsar surveys. Our work here represents a small case study which we hope can be of use in the planning and implementation of the next generation of pulsar surveys on telescopes including MeerKAT, the SKA, and on FAST itself. Rather than the typical discussion surrounding large data volumes and the efficient selection of pulsar candidates, we have demonstrated some of the advantages and challenges associated with following up the pulsar discoveries of one telescope with another which is far less sensitive, a scenario which may repeat itself as the next generation of pulsar surveys gets underway. FAST itself has already commenced observations as part of its CRAFTS survey project, and we can expect the steady production of increasing numbers of new pulsar discoveries from FAST, and their follow-up on Parkes, for many years to come.

\section*{Acknowledgements}
 This work is supported by the  National Key R\&D Program of China No. 2017YFA0402600, CAS Strategic Priority Research Program No. XDB23000000, and by the National Natural Science Foundation of China under Grant No. 11690024, 11725313, 11743002, and 11873067. The Parkes Observatory is part of the Australia Telescope National Facility which is funded by the Australian Government for operation as a National Facility managed by the Commonwealth Scientific and Industrial Research Organisation (CSIRO). This work made use of data from the Five-hundred-meter Aperture Spherical radio Telescope (FAST). FAST is a Chinese national mega-science facility, built and operated by the National Astronomical Observatories, Chinese Academy of Sciences (NAOC). We appreciate the efforts of all those in the FAST Collaboration for their support and assistance during these observations. The authors wish to thank Lawrence Toomey for his data management assistance, Simon Johnston for his insight into RVM fitting and analysis techniques, and Jacinta Raheb-Mol and Maggie Lockwood for their assistance with data reduction. QZ is supported by the National Natural Science Foundation of China (U1731218) and the Science and Technology Fund of Guizhou Province ((2016)-4008, (2017)5726-37). ZP is supported by National Natural Science Foundation of China under Grant No. 11703047, 11773041, U1631132 and U1831131. Both ZP and YY are supported by the CAS `Light of West China' Program. WZ is supported by the CAS Pioneer Hundred Talents Program. JW is supported by the Youth Innovation Promotion  Association of Chinese Academy of Sciences.

\section{Author contributions}

AC, LZ, CCM, MY, SW and JBW carried out most of the observations for this project with the Parkes telescope. MC, DJC and MK supported similar observations with the Effelsberg radio telescope. AC and GJC processed the majority of the data sets.  The remaining authors provided input on the paper and/or were involved in the discovery and selection of candidates from the original FAST survey.




\bibliographystyle{mnras}
\bibliography{fast_parkes} 








\bsp	
\label{lastpage}
\end{document}